# Colloidal 2D $WS_2$ nanosheets exhibiting high-purity single photon emission

Markus Xemaire,[a#] Jakob Keck,[a#] Marco Kögel,[b,c] Marc Brecht,[a,d] Jannik C. Meyer,[b,c,e] Alfred J. Meixner,[a,e] Jannika Lauth[a,e,f,g*].

a. Institute of Physical and Theoretical Chemistry, Eberhard Karls University of Tübingen, Auf der Morgenstelle 18, D-72076 Tübingen, Germany
b. Natural and Medical Sciences Institute (NMI) at the Eberhard Karls University of Tübingen, Markwiesenstraße 55, D-72770 Reutlingen, Germany
c. Institute of Applied Physics, Eberhard Karls University of Tübingen, Auf der Morgenstelle 10, D-72076 Tübingen, Germany
d. Center for Process Analysis and Technology (PA&T), School of Life Sciences and Reutlingen Research Institute (RRI), Reutlingen University, Alte Burgstraße 150, D-72762, Reutlingen, Germany
e. Center for Light-Matter-Interaction, Sensors, and Analytics (LISA+), Eberhard Karls University of Tübingen, Auf der Morgenstelle 15, D-72076 Tübingen, Germany
f. Laboratory for Nano and Quantum Engineering (LNQE), Leibniz University Hannover Schneiderberg 39, D-30167 Hannover
g. Cluster of Excellence PhoenixD (Photonics, Optics, and Engineering – Innovation Across Dimensions), Welfengarten 1A, D-30167 Hannover, Germany

**ABSTRACT:** The search for quantum emitters, delivering single photons "on demand" at cryogenic and elevated temperatures for photonic technologies is ever progressing. For example, μm-sized $WS_2$ monolayers exhibit single photon emission by elaborated local defect or strain engineering. Here, we show that wet-chemically synthesized monolayered two-dimensional (2D) $WS_2$ nanosheets (NSs) with lateral sizes of $14 \pm 6\ nm$ readily emit single photons from exciton recombination without the need for substrate-induced patterning, strain/defect engineering or spectral filtering. Single particle spectroscopy of 70 individual NSs shows photoluminescence from excitons (2.02 eV), trions (1.94 eV) and defect-localized states (1.85 eV) with fast emission ($\tau_1 < 0.2\ ns,\ \tau_2 = 3.2\ ns$), allowing highly frequent excitation cycles. Pulsed at 80 MHz, colloidal $WS_2$ NSs exhibit limited spectral diffusion, no blinking and distinct photon antibunching ($g^2(0) = 0.07$ at 5 K), thereby tackling challenges of established colloidal quantum dot systems (blinking, longer emission cycles), while retaining high single photon purity as key quality of deterministic quantum emitters.

Colloidal nanocrystals (NCs) have gained increasing attention in the pursuit of solution-processable, photostable single photon emitters (SPEs) for photonic quantum technologies including quantum communication.[1, 2] E.g., the generation of single photons "on demand" and in high purity are highly sought for properties to encode information for secure photonic quantum communication.[3]
Multiple zero-dimensional (0D) quantum dot (QD) systems (CdSe/CdS,[4, 5] $CsPbBr_3$,[6-8] PbS/CdS,[9, 10] InP[11] and InAs[12]) and more recently 2D quantum wells[13] have shown to be efficient SPEs, that continuously improve in photoluminescence (PL) quantum yield and single photon purity (SPP).[14] Common motifs in 2D NC-based SPEs are high exciton binding energies, due to confinement in the vertical and/or lateral dimension of the structures, which leads to increased Coulomb interaction of the electron and hole and decreased dielectric screening (in the surrounding medium compared to the NC lattice) of the charge carriers.[15] As a result of the increased exciton-exciton or exciton-charge carrier interaction, nonradiative Auger recombination is increasingly more likely in confined systems[16] and can outcompete radiative recombination.[17] This mechanism can, upon multiple exciton generation, effectively prohibit emission of multiple photons, therefore allowing only SPE.[18] Increased quantum confinement led to high SPP of up to 98% in colloidal 0D $CsPbI_3$ NCs by strongly reducing the biexcitonic quantum yield.[7] While the photostability of NCs is often superior to organic dyes at room temperature, inhomogeneous broadening, spectral diffusion and blinking or "off-times" pose a challenge.[3]

Transition metal dichalcogenides (TMDCs), in particular $WS_2$ monolayers (MLs), are widely studied for their interesting photophysical properties including strong exciton binding energies (0.32 eV),[19] spin polarization,[20] fast radiative recombination (ranging from picoseconds up to ~10 ns)[21-23] and blinking resistance.[24, 25] This combination of properties is highly desirable for advancing the field of colloidal SPEs.

In laterally extended 2D TMDCs SPE has been observed especially near the edges, where point defects such as sulfur vacancies lead to localized excitons and associated SPE.[25] In an effort to control and scale this topography-driven SPE, holes have been deliberately etched[26] or $H_2$-filled domes[27] and $SiO_2$ nanopillars[28, 29] have been used to introduce a predictable amount of strain over a large surface.

Typically, TMDCs are obtained either by mechanical,[30] chemical[31] or liquid exfoliation[32] of bulk crystals or by direct growth on substrates from the gas phase by chemical vapor deposition (CVD).[33, 34]

Distinct from the mentioned techniques is our approach to localize the emission of TMDCs to small emission centers hosting SPE, which is achieved by wet-chemical synthesis of small colloidal TMDCs. A growing number of syntheses have been developed from solution, gaining control over the crystal phase,[35, 36] average layer number,[37] lateral size,[38] and elemental composition.[39] While several studies report on the charge carrier dynamics of colloidal TMDCs with properties comparable to exfoliated samples,[40, 41] the lateral layer size (especially in ML-rich samples) seems (so far) limited to a tunability between $\sim 4 - 40$ nm.[38, 42, 43] Recently, we have studied the excited state dynamics of colloidal $MoS_2$ and $WS_2$ with different lateral sizes (from 3.7 nm to 21.2 nm)[42] as well as alloyed $Mo_{1-x}W_xS_2$ $(0 \leq x \leq 1)$.[44] Density functional theory (DFT) and transient absorption spectroscopy (TAS) revealed a hole-trapping mechanism in colloidal $MoS_2$ and $WS_2$, which results in a sub-picosecond decay of excitons. The responsible mid-gap states are found to be bright and located at the edges of the 2D nanostructures.[42] This possible origin for localized emission is especially important in laterally smaller colloidal $WS_2$ MLs and intrinsic to the NSs in contrast to i.e. substrate induced strain in µm-sized $WX_2$ (X = S, Se) flakes, from which localized emission has been observed previously.[27-29] Darlington *et al*. calculated and probed the local confinement potential exerted by a nanobubble templated substrate (30-140 nm bubble size), implying that “quantum dot like states” responsible for SPE originate from excitons localized to ~10 nm sized wrinkles around each nanobubble.[45] Although the radius of the here examined NSs (~7 nm) exceeds the small exciton Bohr radius of monolayer $WS_2$ (2.6 nm),[46] lateral size does still play a role. For example, liquid exfoliated TMDCs exhibited optoelectronic properties distinct from the central region in a ~8 nm thick so-called “edge region” around each NS.[47] Taking the exciton diffusion in the 2D plane into account (20 nm mean path length),[48] a “soft” confinement was shown to accelerate the Auger rate for lateral sizes even above ten times the exciton Bohr radius.[49]

The $WS_2$ NSs in this work exhibit ~five times the size of the exciton Bohr radius. A “soft” confinement effect could therefore play an additional role in the generation of single photons akin to the mechanism described by Klimov et al for CdSe and PbSe QDs.[15, 16]

Here we show SPE from wet-chemically synthesized $WS_2$ NSs with a high SPP of 93% (second-order correlation function ($g^2(0) = 0.07$)) at 5 K and antibunching up to at least 110 K (SPP ~80%). By evaluating 70 individual NSs, different contributions from neutral excitons ($X_A$), negative trions ($T_-$), spin and/or momentum forbidden transitions and defect localized emission are found and are in excellent agreement with literature on comparable exfoliated and CVD-grown samples.[22, 23, 50, 51] Time-correlated single photon counting (TCSPC) shows a fast biexponential decay ($\tau_1 < 0.2\,ns$, $\tau_2 = 3.2\,ns \pm 20\,ps$), allowing emission of single photons at rates of 80 MHz. These results demonstrate the highly promising SPE properties of colloidal $WS_2$ NSs as deterministic single photon source.[52]

Colloidal $WS_2$ NSs are synthesized by dropwise addition of $WCl_6$ dissolved in oleylamine (OlAm, 10 ml) to a hot (320°C) sulfur precursor solution at a rate of 10 ml/h (see Methods S6 for detailed description). Figure 1a shows an example of an individual ML of semiconducting 2H-phase $WS_2$, as is evident by the first and second order signals of the hexagonal crystal structure (yellow arrows) spaced 0.23 and 0.13 $nm^{-1}$, in agreement with the lattice parameter of $WS_2$ ($a = 3.15$ Å).[53, 54] Only one orientation is visible for this single $WS_2$ NS (for comparison see Fig. S1 with an overview image containing several $WS_2$ NSs including multiple orientations and flat lying MLs). Few-layers are identified by an interlayer distance of 0.3 nm,[54] while ligand separation keeps the majority of monolayered NSs from direct stacking, thereby preserving their optical properties (PL,[55-57] SHG,[56] position of spectral features).[37, 39, 40, 42, 55, 56] The NSs have an average lateral size of 14 ± 6 nm as measured individually from transmission electron microscopy (TEM) images (see Fig. 1b for size distribution). The size is measured along the longest side of each NS. Taking a rectangular aspect ratio into account, the average area is ~140 $nm^2$. While all colloidal synthesis protocols for TMDCs yield an ensemble containing few-layers or small amounts of the metallic 1T-phase,[35, 36, 38] we previously described a synthesis protocol for predominantly monolayered and semiconducting NSs.[39, 56]

Analyzing the absorption of a colloidal solution of $WS_2$ NSs in Fig. 1c, four excitonic transitions (A: 2.02 eV/614 nm, B: 2.40 eV/517 nm, C: 2.85 eV/435 nm, D: 3.40 eV/365 nm) are observed in good agreement with values reported for ML samples.[23, 40, 51, 55, 56] The A and B excitons in ML $WS_2$ are separated by strong spin-orbit coupling (~0.4 eV) and reside both at the K-point of the Brillouin zone as direct hill-valley transitions. While PL from the B exciton can be observed in $WS_2$,[58] the lowest energy transition, the A-exciton, dominates the PL.

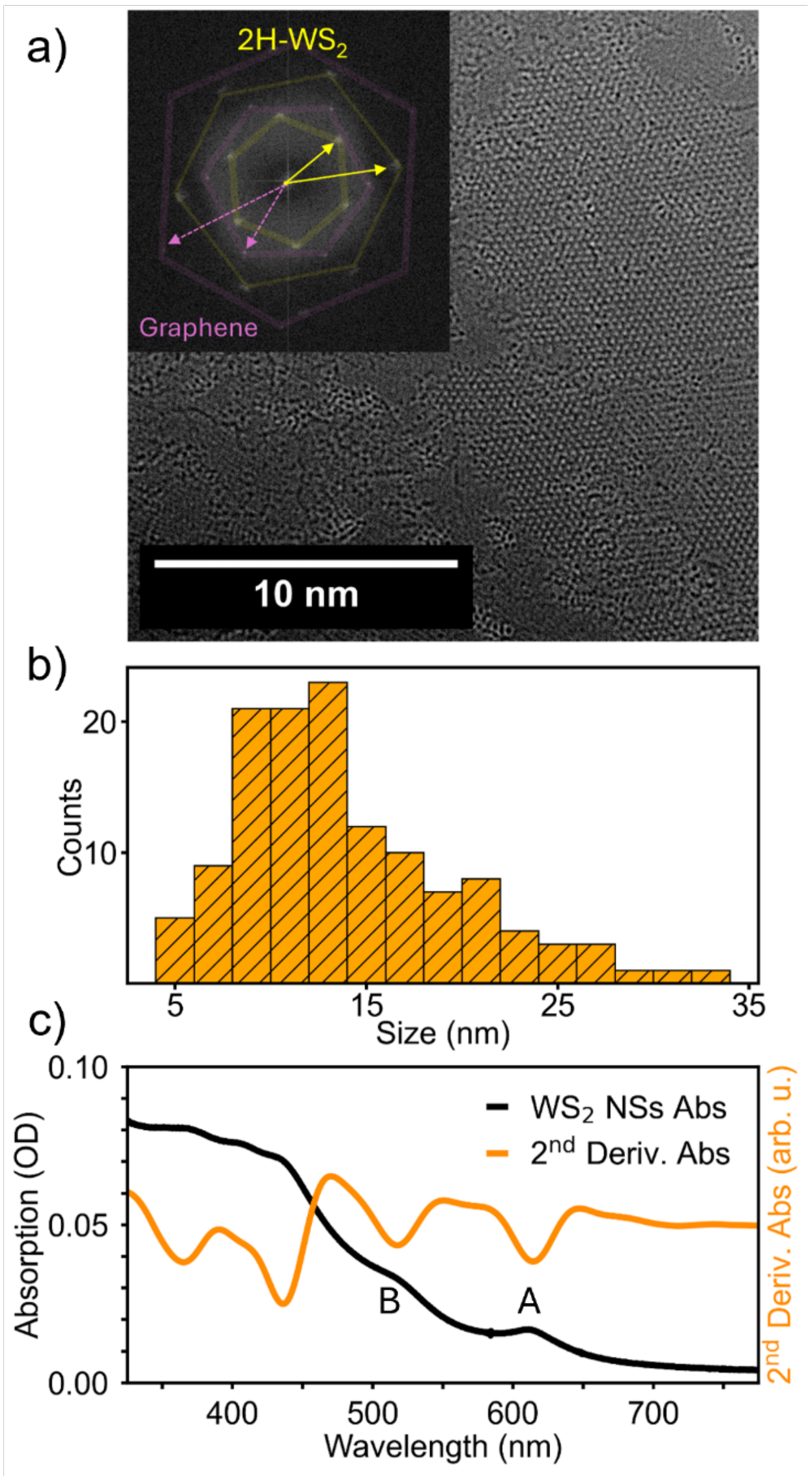

Fig. 1. a) HR-TEM image showing a single $WS_2$ NS with ~10 x 14 nm (140 nm$^2$) size, the FFT inset shows the first and second order signals for the hexagonal crystal structure of 2H-phase $WS_2$ in a single orientation (yellow arrows), as well as for the underlying graphene layer (purple arrows), b) size distribution of $WS_2$ NSs (averaging 14 ± 6 nm) obtained by measuring 130 NSs from TEM images, c) absorption spectrum of colloidal $WS_2$ NSs and associated second derivative plotted for better visibility of the A, B, C and D excitonic features.

Fig. 2a) contains a series of spectra from a single $WS_2$ NS, each integrated for 1 s with an excitation pulse rate of 80 MHz. The sample was cooled to 4.8 K in a liquid helium bath cryostat (see Methods S3 and Fig. S2 for a detailed description of the setup and experiment). The distribution of sizes observed in the ensemble (Fig. 1b, c) is missing, reducing the inhomogeneous line-broadening in the spectra.[59] Reduced phonon interaction at 4.8 K leads to additional narrowing[60] and to a decrease in the lattice expansion of the $WS_2$ NS, as well as reduced exciton-phonon interaction and consequently a decreased red-shift.[61] Another effect on the position of PL features to keep in mind is pre-selection. $WS_2$ exhibits a direct band gap in MLs only, few-layers contribute marginally to the PL of a mixture of mono- and few-layers.[59] For example, a single NS spectrum can be slightly blue-shifted compared to the respective absorption of the ensemble because the ML content of the photoluminescent subset is higher.[56]

The measurement of single NSs allows for the characterization of variations between individual emitters[62] and dynamic fluctuations such as spectral diffusion[63, 64] or blinking.[65, 66] The observed NS exhibits stable emission without blinking for at least 180 s (Fig. 2a). Further example series show good stability and blinking-free emission for most of $WS_2$ NSs studied at ~5 K (Fig. S3 a-g). At elevated temperatures, NSs tend to be less stable, which is why shorter series were collected (see Fig. S3 h-l) to obtain PL lifetimes and subsequent antibunching experiments, that are discussed below. Between 112 and 122 s (Fig. 2a), collective spectral diffusion of all features is visible. This means that temporal changes of the dielectric environment, e.g. the conformation or the ligand shell of the NS affect all optical transitions equally, producing a simultaneous shift across all features (Fig. 2b). This behavior is only observed when a single NS is probed and serves as additional proof that diluting the colloidal $WS_2$ solution ($\times 10^{-6}$) yields single widely spaced NSs, a necessary condition for the following experiments (see Fig. S4 for spatially resolved scan images with widely spaced $WS_2$ NSs).

PL spectra of TMDCs typically show multiple contributions from neutral excitons ($X_A$), negative trions ($T_-$) or defect localized emission, even at room temperature.[50, 55, 67] In earlier measurements on colloidal $WS_2$ NSs at room temperature, we found these spectral features at $X_A = 2.02\ eV$, $T_- = 1.94\ eV$ and $X_D = 1.85\ eV$ in good accordance with the positions obtained here (see Fig. 4 for 295 K spectrum) and in literature.[50, 56, 68] Even with the undiluted concentration in these earlier measurements, spatial fluctuations in contribution from excitons, trions and defect emission became apparent, underlining the need for single NS spectroscopy.[56]

At 5 K, the features of single $WS_2$ NSs shift blue ($X_A = 2.08\ eV$, $T_- = 2.05\ eV$, $X_D = 1.91\ eV$), while additional features of spin and/or momentum forbidden transitions (summarized in Fig. S5) arise and are reported to dominate the spectra at low temperatures.[22, 23] In colloidal $WS_2$ NSs the highest energy neutral exciton transition ($X_A$) stays the dominant feature even at 5 K. By aligning all spectra recorded at 5 K to the zero-phonon line (ZPL) of the highest energy transition ($X_A$), a series of distinct features is obtained, that are assigned according to literature as follows (see Fig. 2c): At -9 and -18 meV relative to the ZPL the first ($r_1$) and second ($r_2$) phonon replica are visible as shoulders of decreasing intensity. Repeating spacings of similar energy (7-9 meV) were reported in exfoliated $WS_2$ monolayers to originate from interlayer breathing modes between the monolayer and the substrate (a silicon wafer).[23] Between -20 to -50 meV, negative trions are reported,[22] while further momentum- and/or spin-forbidden transitions are probed up to -100 meV.[22, 51] Although few studies report features beyond a shift of 100 meV relative to the ZPL, cathodoluminescence studies on exfoliated $WS_2$ showed deeper emissive trap levels between -130 and -190 meV.[69] The origin of trap states in $WS_2$ are known to be manifold. For example, abundant sulfur vacancies, both on the basal plane[70] and the edge region,[59] introduce mid-gap states. We recently combined DFT with TAS and found that even saturated edges can lead to emissive mid-gap states close to the valence band, especially impacting laterally small TMDCs.[42] We therefore assume both, sulfur vacancies and/or edge localized hole traps are possible origins for emissive transitions beyond -100 meV relative to the ZPL.

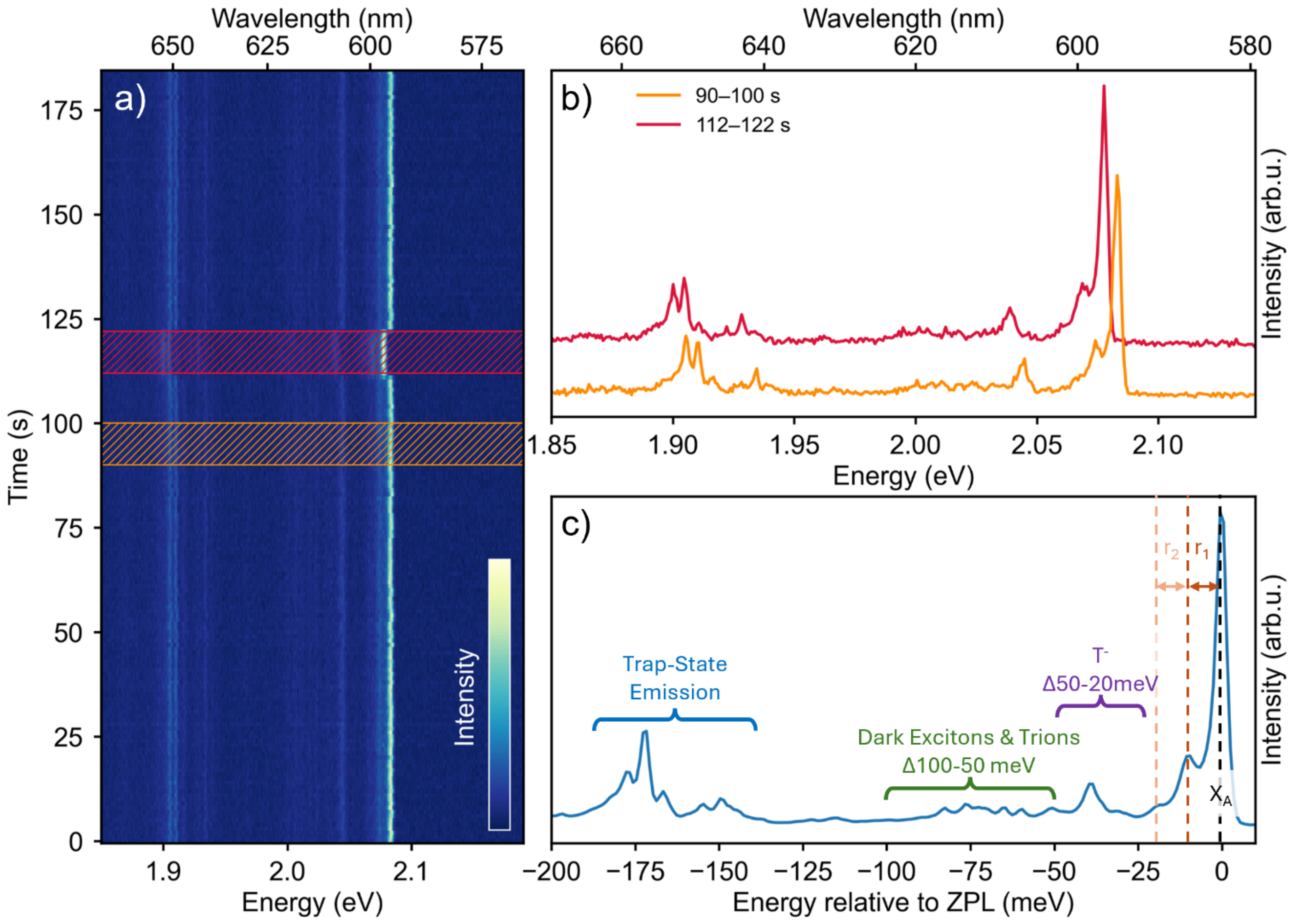

Fig. 2. a) Sequence of 180 PL spectra recorded from a single $WS_2$ NS at 5 K, with 1 s integration time per spectrum. The sequence shows the zero phonon (ZPL) line at 2.08 eV with additional narrow features at lower energies. Small, discrete reversible spectral jumps, and one larger reversible jump of 5.5 meV between spectrum 112 and 122 which are synchronous for all lines are a typical sign for a single NS. b) PL spectra averaged over the red and orange marked range, show this shift is synchronous for all spectral features, c) averaged spectrum of all 5 K measurements aligned to the ZPL, showing discrete features shifted relative to the ZPL, which are attributed to bright transitions including neutral excitons ($X_A$, 0 meV), negative trions ($T_{-}$, 50-20 meV) as well as a variety of spin and/or momentum forbidden dark transitions (100-50 meV) and emissions from trap states (190-130 meV).

In a Hanbury Brown and Twiss setup,[71] a 50/50 splitter divided the emitted photons to be detected by one of two avalanche photodiodes (APDs) (see Methods S3, Fig. S2). By calculating the second-order detector-correlation function at the time of excitation ($g^2(0)$), the single photon purity (SPP) of the sample is obtained. While multiple photon sources (e.g. thermal light) give $g^2(0)$ values of $>1$, $g^2(0) < 1$ are obtained only from quantum emitters, where one photon is emitted at a time.[3] When observing an ensemble, the following equation applies: $g^2(0) = 1 - 1/N$, where $N$ is the number of quantum emitters of the observed ensemble. Second-order correlations of $1 > g^2(0) > 0.5$ are obtained from multiple single photon sources, while $g^2(0) < 0.5$ is only possible to originate from one individual single photon source (N=1), proving antibunched emission that is considered a typical property of a single photon light source.[72, 73] Since we are observing single NSs, as was established above, we expect $g^2(0) < 0.5$, while a value of 0 would be a perfectly pure single photon emitter. Figure 3a shows the $g^2(0)$ of a single colloidal $WS_2$ NS excited at 520 nm with an 80 MHz pulse rate. A value of $g^2(0) = 0.07 \pm 0.03$ is calculated, corresponding to a high SPP of 90-96% including errors (see Methods S7 for a detailed calculation). The TCSPC histogram (see Fig. 3b), reveals a fast biexponential radiative decay, allowing SPE with highly frequent excitation cycles. The first decay component lies below the IRF fall time of 200 ps, which is taken as an approximate minimal measurable decay time of the setup ($\tau_1 < 0.2\, ns$, $\tau_2 = 3.2\, ns \pm 20\, ps$). These values are (although not resolving picosecond scale phonon interactions)[44, 60] well within the range of reported $WS_2$ PL lifetimes ranging from picoseconds up to ~10 ns depending on layer number[21, 23, 74] or surface treatment.[75] In earlier studies with colloidal samples, TAS showed comparable signal decays that were associated with radiative recombination processes.[42, 44]

To underpin the potential of colloidal $WS_2$ NSs for SPE at elevated temperatures, antibunching of $WS_2$ NSs is shown at 85 and 110 K respectively (see Fig. 3c). Please note that the increased noise widens the standard deviation and causes the $g^2(0)$ values to spread from 0-0.1 for 85 K and 0.12-0.3 for 110 K respectively. At both temperatures, the emission of the measured NSs is still antibunched ($g^2(0) < 0.5$). For now, it is unclear if the SPP is reduced at elevated temperatures or if increased noise is the main factor limiting the $g^2(0)$ measurement. Current experiments therefore include the synthesis of colloidal $WS_2$ NSs with increased quantum yield for measuring SPE up to room temperature.

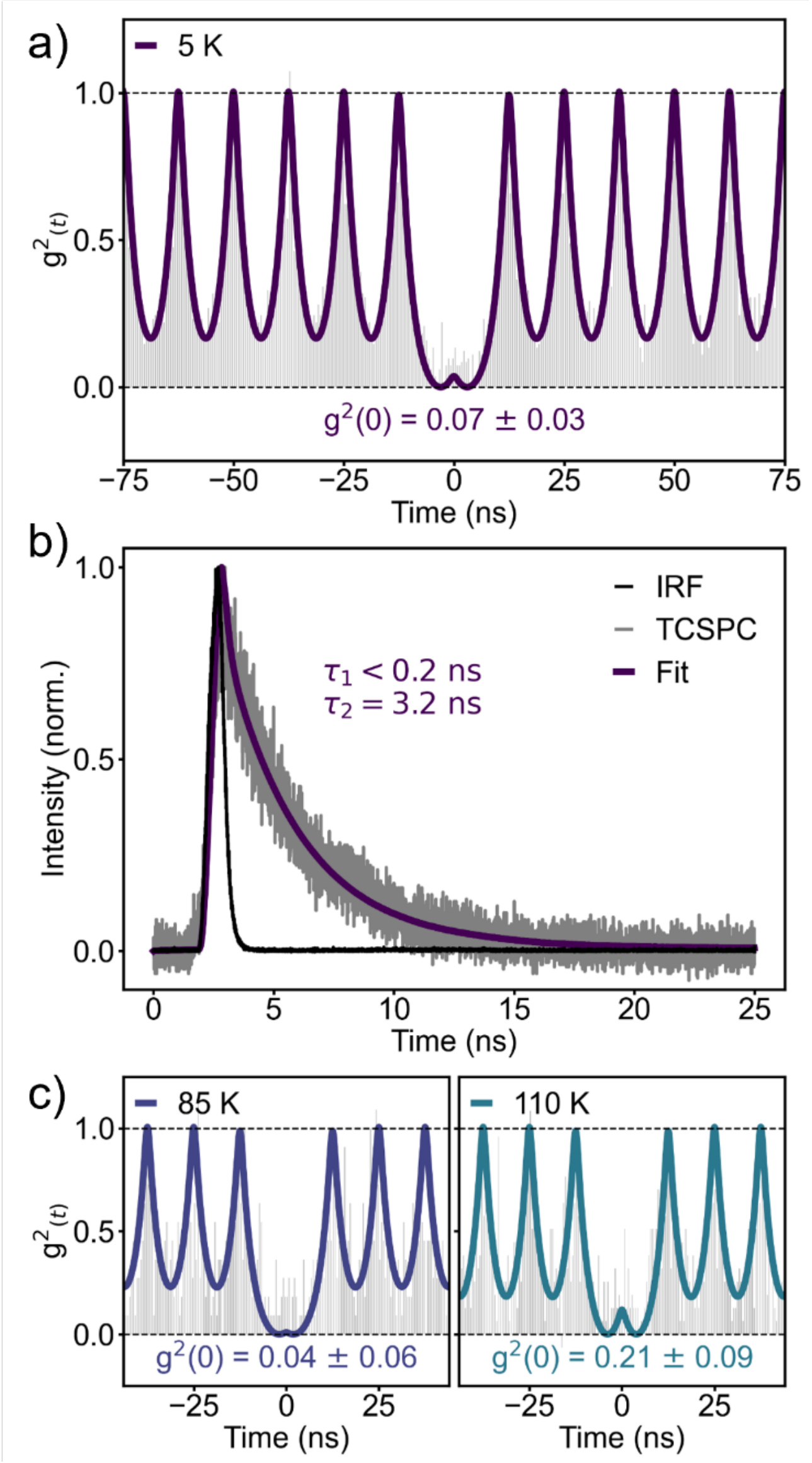

Fig. 3. a) Second-order photon correlation function $g^2(\tau)$ of a colloidal $WS_2$ NS at 5 K showing a high SPP of 93% ($g^2(0) = 0.07 \pm 0.03$), b) TCSPC histogram of a $WS_2$ NS at 5 K exhibiting biexponential decay dynamics ($\tau_1 < 0.2\,ns$, $\tau_2 = 3.2\,ns \pm 20\,ps$), c) antibunching measurements at 85 K and 110 K, with lower PL intensity at elevated temperatures leading to increased noise.

In total ~70 single $WS_2$ NSs were characterized with different methods (TCSPC, PL and antibunching). This allows for the investigation of temperature dependent trends in the emission of colloidal $WS_2$. Fig. 4a shows spectra for each of the temperature steps (5-295 K), qualitatively visualizing the apparent red-shift of the PL and spectral broadening with increasing temperature. A gradual intensity loss due to phonon-mediated nonradiative recombination[76] is observed by the increased noise at elevated temperatures (note that PL intensity loss caused by exciton dissociation is highly unlikely up to room temperature ($k_B \cdot 295\,K = 0.03\,eV$), due to the high exciton binding energy of $WS_2$ MLs (0.32 eV).[19]

Differences between individual $WS_2$ NSs can be quite substantial, since environmental influences (e.g. polymer matrix, underlying Si-wafer), lateral size, edge configuration and strain may vary from NS to NS.[59] To quantify the peak position of the neutral exciton $X_A$ in dependence of temperature (see Fig. 4b), box-whisker plots are applied. The sample size ($n = 20$, for 5 K) varies for each temperature step. All data points lie between the whiskers, while the box accounts for 50% of data and the median is marked as white line within the box. As phonon interactions increase with temperature, a red-shifting median peak position (60 meV) is observed.[50, 59] Fitting the temperature dependent peak positions ($E_g(T)$) to a Varshni equation $E_g(T) = E_g(0) - \alpha T^2/(T+\beta)$[77] and approximating 2.08 eV as $E_g(0)$, an empirical factor describing the lattice dilation and therefore the narrowing of the band gap with increasing temperature ($\alpha = 3.8 \times 10^{-4} eV/K$, see Fig. S6 a) is found for $\beta = 200\,K$ (empirical factor related to the Debye temperature). Even with six temperature steps, the resulting $\alpha$ is in good agreement with studies on single $WS_2$ MLs obtained from exfoliation.[50] A similar evaluation was performed for the full width at half maxima (FWHM, see Fig. 4c). At 5 K, the observed FWHM of the $X_A$ feature varies from 2.87 – 4.7 meV and is narrow compared to free excitons (~2-4 and 10 meV)[78, 79] in laterally extended $WS_2$ and $WSe_2$ respectively, while the minimal linewidth (2.87 meV see Fig. S7) is limited by the spectral resolution of the setup (0.8 nm or 2.87 meV at 2.08 eV, see Methods S3). Differences in linewidth between individual NSs may result from intrinsic effects including defects in the crystal structure or extrinsic effects of the environment i.e. dielectric fluctuations in the medium/matrix or dangling bonds of potential underlying substrate, when comparing with encapsulated samples.[80] Excluding the FWHM at 5 K, a nearly linear scaling broadening is observed ($4 \times 10^{-4} eV/K$, see Fig. S6 b) between 85 and 295 K, as is expected, while a visible curvature is described between 5 and 100 K with higher sampling rate.[60, 81]

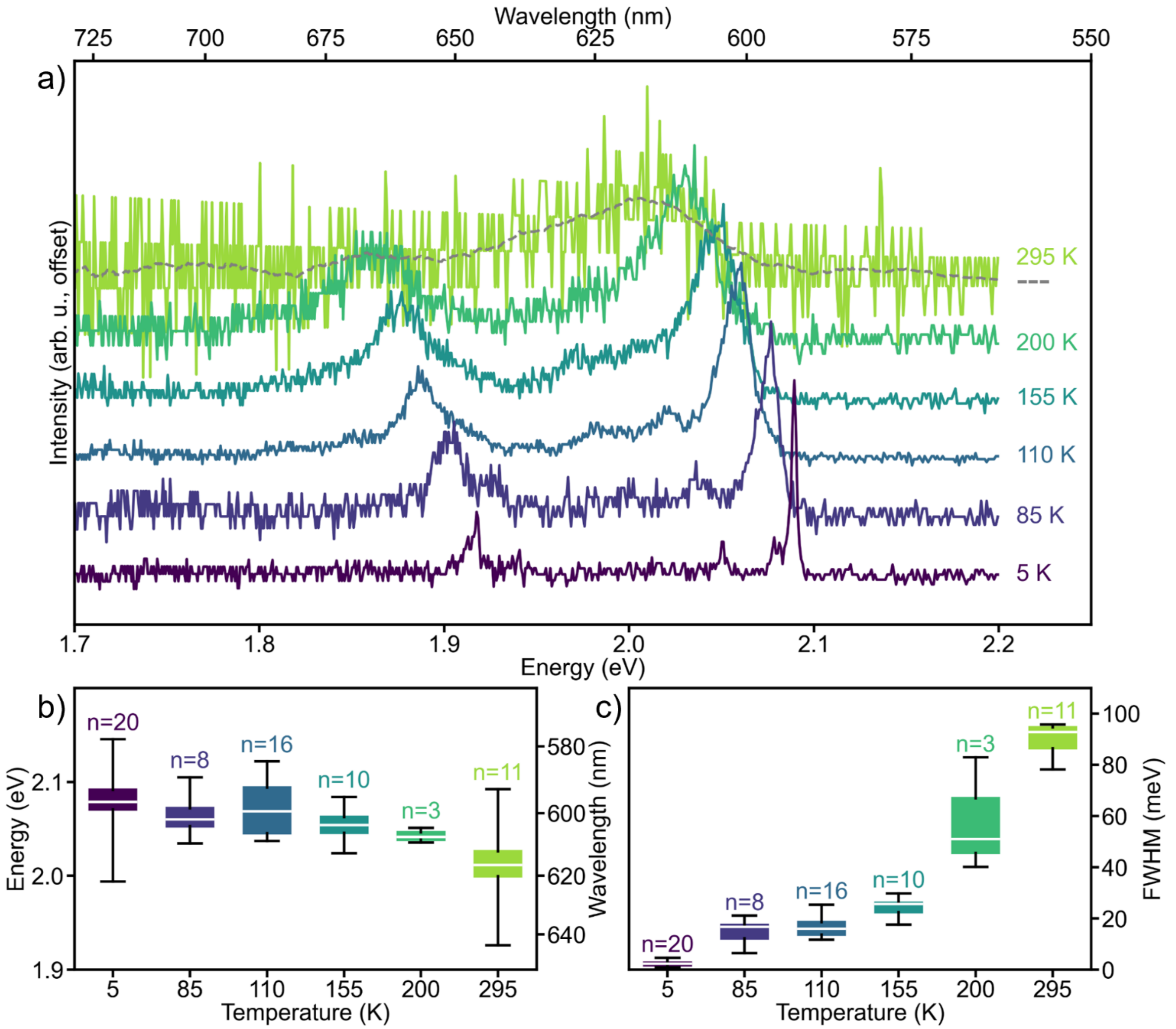

Fig. 4. a) PL spectra (1 s integration time) of individual $WS_2$ NSs as a function of temperature show a pronounced red-shift and line broadening towards higher temperatures. The gray dashed line serves to visualize the noisier spectrum at 295 K, box whisker plots of temperature-dependent ZPL position b) and FWHM c). All datapoints lie between the whiskers, while the box contains a median value (white line) and 25% of data above and below the median, sample size (n) is displayed for each temperature, showing a clear trend with increasing temperature despite the variation of individual NSs for both ZPL position (red-shift of 60 meV) and FWHM (broadening of $4 \times 10^{-4} eV/K$).

In conclusion, we have shown that wet-chemically synthesized $WS_2$ NSs of ~140 nm$^2$ lateral size exhibit stable and efficient SPE with a high SPP of 93% ($g^2(0) = 0.07 \pm 0.03$) at 5 K and antibunching up to at least 110 K (SPP ~80%), without posterior treatment like substrate patterning or strain/defect engineering. Due to the strong excitonic interactions previously probed by TAS in these highly confined materials, we infer that an additional "soft" confinement in the 2D-plane of the NSs (contrary to µm-sized $WS_2$ flakes) could effectively limit the emission to single photons as an intrinsic mechanism akin to CdX, PbX (X=S,Se) or $CsPbY_3$ (Y= Cl, Br, I) QDs.

By evaluating ~70 individual $WS_2$ NSs, complex spectral features with contribution from neutral excitons, negative trions and trap-state emission are described. The measurements reveal the individual dynamic fluctuation of each NS and show the majority of NSs to continuously emit with little spectral diffusion (especially at 5 K) and without blinking. The minimal linewidth of the neutral exciton at 5 K is limited by the spectral resolution of the spectrograph to 2.87 meV. TCSPC shows a fast biexponential decay ($\tau_1 < 0.2\,ns$, $\tau_2 = 3.2\,ns \pm 20\,ps$), that allows for frequent re-excitation of the system, ultimately realizing 93% SPP at 80 MHz pulse rate without spectral filtering. Our work represents an important step toward wet-chemical SPEs with "on-demand", fast, stable and high purity intrinsic SPE.

Stability of All Inorganic Perovskite Nanocrystals for Single Photon Emission. *Advanced Optical Materials* **2025**, *13* (14), 2403441. DOI: https://doi.org/10.1002/adom.202403441.

## ASSOCIATED CONTENT

Supporting information including: synthesis protocol, experimental setup, overview TEM images, antibunching at elevated temperatures, exemplary calculations and further spectral data mentioned in the main manuscript.

## AUTHOR INFORMATION

### Corresponding Author

*Jannika Lauth
Email: jannika.lauth@uni-tuebingen.de

### Author Contributions

All authors have given approval to the final version of the manuscript. M. X. and J. K. contributed equally to this work. Conceptualization: M. X., J. K and J. L.; synthesis: M. X., investigation: M. X., J. K, M. K., data processing: M. X., J. K. visualization and formal analysis: M. X.; methodology: M. X., J. K.; resources: A. J. M., J. C. M. and J. L.; writing original draft: M. X.; review and editing: M. X., J. K., A. J. M., J. C. M. and J. L.

## ACKNOWLEDGMENT

We gratefully acknowledge financial support by the Deutsche Forschungsgemeinschaft (DFG, German Research Foundation) projects LA 3649/4-1, ME3133/6-1 and INST37/1138-1 as well as the Germany's Excellence Strategy within the Cluster of Excellence PhoenixD (EXC 2122, Project ID 390833453).

## REFERENCES

(1) Kagan, C. R.; Bassett, L. C.; Murray, C. B.; Thompson, S. M. Colloidal Quantum Dots as Platforms for Quantum Information Science. *Chemical Reviews* **2020**, *121* (5), 3186-3233. DOI: 10.1021/acs.chemrev.0c00831.

(2) Senellart, P.; Solomon, G.; White, A. High-performance semiconductor quantum-dot single-photon sources. *Nature Nanotechnology* **2017**, *12* (11), 1026-1039. DOI: 10.1038/nnano.2017.218.

(3) Lounis, B.; Orrit, M. Single-photon sources. *Reports on Progress in Physics* **2005**, *68* (5), 1129. DOI: 10.1088/0034-4885/68/5/R04.

(4) Morozov, S.; Pensa, E. L.; Khan, A. H.; Polovitsyn, A.; Cortés, E.; Maier, S. A.; Vezzoli, S.; Moreels, I.; Sapienza, R. Electrical control of single-photon emission in highly charged individual colloidal quantum dots. *Science Advances* **2020**, *6* (38), eabb1821. DOI: doi:10.1126/sciadv.abb1821.

(5) Nguyen, H. A.; Sharp, D.; Fröch, J. E.; Cai, Y.-Y.; Wu, S.; Monahan, M.; Munley, C.; Manna, A.; Majumdar, A.; Kagan, C. R.; et al. Deterministic Quantum Light Arrays from Giant Silica-Shelled Quantum Dots. *ACS Applied Materials & Interfaces* **2023**, *15* (3), 4294-4302. DOI: 10.1021/acsami.2c18475.

(6) Utzat, H.; Sun, W.; Kaplan, A. E. K.; Krieg, F.; Ginterseder, M.; Spokoyny, B.; Klein, N. D.; Shulenberger, K. E.; Perkinson, C. F.; Kovalenko, M. V.; et al. Coherent single-photon emission from colloidal lead halide perovskite quantum dots. *Science* **2019**, *363* (6431), 1068-1072. DOI: doi:10.1126/science.aau7392.

(7) Zhu, C.; Marczak, M.; Feld, L.; Boehme, S. C.; Bernasconi, C.; Moskalenko, A.; Cherniukh, I.; Dirin, D.; Bodnarchuk, M. I.; Kovalenko, M. V.; et al. Room-Temperature, Highly Pure Single-Photon Sources from All-Inorganic Lead Halide Perovskite Quantum Dots. *Nano Letters* **2022**, *22* (9), 3751-3760. DOI: 10.1021/acs.nanolett.2c00756.

(8) Gorji, S.; Pashaei Adl, H.; Gualdrón-Reyes, A. F.; Sánchez Sánchez, A.; Sánchez Alarcón, R. I.; Pareja-Rivera, C.; Mora-Seró, I.; Martínez Pastor, J. P.; Muñoz Matutano, G. Enhanced Optical

(9) Hu, Z.; Kim, Y.; Krishnamurthy, S.; Avdeev, I. D.; Nestoklon, M. O.; Singh, A.; Malko, A. V.; Goupalov, S. V.; Hollingsworth, J. A.; Htoon, H. Intrinsic Exciton Photophysics of PbS Quantum Dots Revealed by Low-Temperature Single Nanocrystal Spectroscopy. *Nano Letters* **2019**, *19* (12), 8519-8525. DOI: 10.1021/acs.nanolett.9b02937.

(10) Krishnamurthy, S.; Singh, A.; Hu, Z.; Blake, A. V.; Kim, Y.; Singh, A.; Dolgopolova, E. A.; Williams, D. J.; Piryatinski, A.; Malko, A. V.; et al. PbS/CdS Quantum Dot Room-Temperature Single-Emitter Spectroscopy Reaches the Telecom O and S Bands via an Engineered Stability. *ACS Nano* **2021**, *15* (1), 575-587. DOI: 10.1021/acsnano.0c05907 (acccessed 10/26/2026).

(11) Proppe, A. H.; Berkinsky, D. B.; Zhu, H.; Šverko, T.; Kaplan, A. E. K.; Horowitz, J. R.; Kim, T.; Chung, H.; Jun, S.; Bawendi, M. G. Highly stable and pure single-photon emission with 250 ps optical coherence times in InP colloidal quantum dots. *Nature Nanotechnology* **2023**, *18* (9), 993-999. DOI: 10.1038/s41565-023-01432-0.

(12) Yang, Y.; Li, J.; Li, Y.; Li, B.; Zhang, J.; Chen, Z.; Xiao, L.; Lai, R.; Zeng, Z.; Zhang, G.; et al. High-Purity Single-Photon Emission in Near-Infrared InAs Colloidal Quantum Dots with Strong Exciton Confinement. *Nano Letters* **2025**, *25* (18), 7595-7602. DOI: 10.1021/acs.nanolett.5c01546.

(13) Yin, T.; Liang, X.; Huang, Y.; Thung, Y. T.; Ni, Z.; Sun, H.; Demir, H. V. Engineering Colloidal Quasi-2D Quantum Wells for High-Performance Room-Temperature Single-Photon Sources. *Journal of the American Chemical Society* **2025**, *147* (38), 34540-34547. DOI: 10.1021/jacs.5c08797.

(14) Tomm, N.; Javadi, A.; Antoniadis, N. O.; Najer, D.; Löbl, M. C.; Korsch, A. R.; Schott, R.; Valentin, S. R.; Wieck, A. D.; Ludwig, A.; et al. A bright and fast source of coherent single photons. *Nature Nanotechnology* **2021**, *16* (4), 399-403. DOI: 10.1038/s41565-020-00831-x.

(15) Klimov, V. I. Mechanisms for Photogeneration and Recombination of Multiexcitons in Semiconductor Nanocrystals: Implications for Lasing and Solar Energy Conversion. *The Journal of Physical Chemistry B* **2006**, *110* (34), 16827-16845. DOI: 10.1021/jp0615959.

(16) Klimov, V. I.; Mikhailovsky, A. A.; McBranch, D. W.; Leatherdale, C. A.; Bawendi, M. G. Quantization of Multiparticle Auger Rates in Semiconductor Quantum Dots. *Science* **2000**, *287* (5455), 1011-1013. DOI: doi:10.1126/science.287.5455.1011.

(17) Park, Y.-S.; Guo, S.; Makarov, N. S.; Klimov, V. I. Room Temperature Single-Photon Emission from Individual Perovskite Quantum Dots. *ACS Nano* **2015**, *9* (10), 10386-10393. DOI: 10.1021/acsnano.5b04584 (acccessed 8/19/2026).

(18) Lounis, B.; Bechtel, H. A.; Gerion, D.; Alivisatos, P.; Moerner, W. E. Photon antibunching in single CdSe/ZnS quantum dot fluorescence. *Chemical Physics Letters* **2000**, *329* (5), 399-404. DOI: https://doi.org/10.1016/S0009-2614(00)01042-3.

(19) Chernikov, A.; Berkelbach, T. C.; Hill, H. M.; Rigosi, A.; Li, Y.; Aslan, B.; Reichman, D. R.; Hybertsen, M. S.; Heinz, T. F. Exciton Binding Energy and Nonhydrogenic Rydberg Series in Monolayer $WS_2$. *Physical Review Letters* **2014**, *113* (7), 076802. DOI: 10.1103/PhysRevLett.113.076802.

(20) Kioseoglou, G.; Korkusinski, M.; Scrace, T.; Hanbicki, A. T.; Currie, M.; Jonker, B. T.; Petrou, A.; Hawrylak, P. Spin effects in $MoS_2$ and $WS_2$ single layers. *physica status solidi (RRL) – Rapid Research Letters* **2016**, *10* (1), 111-119. DOI: https://doi.org/10.1002/pssr.201510291 (acccessed 2026/06/15).

(21) Liu, Y.; Hu, X.; Wang, T.; Liu, D. Reduced Binding Energy and Layer-Dependent Exciton Dynamics in Monolayer and Multilayer $WS_2$. *ACS Nano* **2019**, *13* (12), 14416-14425. DOI: 10.1021/acsnano.9b08004.

(22) Bao, D.; del Águila, A. G.; Thu Ha Do, T.; Liu, S.; Pei, J.; Xiong, Q. Probing momentum-indirect excitons by near-resonance photoluminescence excitation spectroscopy in $WS_2$ monolayer. *2D Materials* **2020**, *7* (3), 031002. DOI: 10.1088/2053-1583/ab817a.

(23) Chand, S. B.; Woods, J. M.; Quan, J.; Mejia, E.; Adhikari, A.; Wang, W.; Taniguchi, T.; Watanabe, K.; Alù, A.; Grosso, G. Phonon-Dressed States of Dark Excitons in Transition Metal Dichalcogenides. *Advanced Optical Materials* **2026**, *14* (13), e03833. DOI: https://doi.org/10.1002/adom.202503833 (acccessed 2026/06/15).

(24) Peimyoo, N.; Shang, J.; Cong, C.; Shen, X.; Wu, X.; Yeow, E. K. L.; Yu, T. Nonblinking, Intense Two-Dimensional Light Emitter: Monolayer $WS_2$ Triangles. *ACS Nano* **2013**, *7* (12), 10985-10994. DOI: 10.1021/nn4046002.

(25) Tonndorf, P.; Schmidt, R.; Schneider, R.; Kern, J.; Buscema, M.; Steele, G. A.; Castellanos-Gomez, A.; van der Zant, H. S. J.; Michaelis de Vasconcellos, S.; Bratschitsch, R. Single-photon emission from localized excitons in an atomically thin semiconductor. *Optica* **2015**, *2* (4), 347-352. DOI: 10.1364/OPTICA.2.000347.

(26) Kumar, S.; Kaczmarczyk, A.; Gerardot, B. D. Strain-Induced Spatial and Spectral Isolation of Quantum Emitters in Mono- and Bilayer $WSe_2$. *Nano Letters* **2015**, *15* (11), 7567-7573. DOI: 10.1021/acs.nanolett.5b03312.

(27) Cianci, S.; Blundo, E.; Tuzi, F.; Pettinari, G.; Olkowska-Pucko, K.; Parmenopoulou, E.; Peeters, D. B. L.; Miriametro, A.; Taniguchi, T.; Watanabe, K.; et al. Spatially Controlled Single Photon Emitters in hBN-Capped $WS_2$ Domes. *Advanced Optical Materials* **2023**, *11* (12), 2202953. DOI: https://doi.org/10.1002/adom.202202953.

(28) Parto, K.; Azzam, S. I.; Banerjee, K.; Moody, G. Defect and strain engineering of monolayer WSe2 enables site-controlled single-photon emission up to 150 K. *Nature Communications* **2021**, *12* (1), 3585. DOI: 10.1038/s41467-021-23709-5.

(29) Wu, W.; Strasser, A.; Xiu, Z.; Ngan, K.; Hoang, T. D.; Liu, S.; Shen, H.; Holtzman, L. N.; Edgar, J. H.; Hone, J. C.; et al. Ultrahigh-Purity Single-Photon Emission from 2D $WSe_2$ via Effective Suppression of Classical Emission. *Nano Letters* **2025**, *25* (29), 11226-11233. DOI: 10.1021/acs.nanolett.5c01718.

(30) Li, H.; Wu, J.; Yin, Z.; Zhang, H. Preparation and Applications of Mechanically Exfoliated Single-Layer and Multilayer $MoS_2$ and $WSe_2$ Nanosheets. *Accounts of Chemical Research* **2014**, *47* (4), 1067-1075. DOI: 10.1021/ar4002312.

(31) Eda, G.; Yamaguchi, H.; Voiry, D.; Fujita, T.; Chen, M.; Chhowalla, M. Photoluminescence from Chemically Exfoliated $MoS_2$. *Nano Letters* **2011**, *11* (12), 5111-5116. DOI: 10.1021/nl201874w.

(32) Coleman, J. N.; Lotya, M.; O'Neill, A.; Bergin, S. D.; King, P. J.; Khan, U.; Young, K.; Gaucher, A.; De, S.; Smith, R. J.; et al. Two-Dimensional Nanosheets Produced by Liquid Exfoliation of Layered Materials. *Science* **2011**, *331* (6017), 568-571. DOI: 10.1126/science.1194975.

(33) Wang, X.; Gong, Y.; Shi, G.; Chow, W. L.; Keyshar, K.; Ye, G.; Vajtai, R.; Lou, J.; Liu, Z.; Ringe, E.; et al. Chemical Vapor Deposition Growth of Crystalline Monolayer $MoSe_2$. *ACS Nano* **2014**, *8* (5), 5125-5131. DOI: 10.1021/nn501175k.

(34) Lee, Y.-H.; Zhang, X.-Q.; Zhang, W.; Chang, M.-T.; Lin, C.-T.; Chang, K.-D.; Yu, Y.-C.; Wang, J. T.-W.; Chang, C.-S.; Li, L.-J.; et al. Synthesis of Large-Area $MoS_2$ Atomic Layers with Chemical Vapor Deposition. *Advanced Materials* **2012**, *24* (17), 2320-2325. DOI: https://doi.org/10.1002/adma.201104798.

(35) Mahler, B.; Hoepfner, V.; Liao, K.; Ozin, G. A. Colloidal Synthesis of 1T-$WS_2$ and 2H-$WS_2$ Nanosheets: Applications for Photocatalytic Hydrogen Evolution. *Journal of the American Chemical Society* **2014**, *136* (40), 14121-14127. DOI: 10.1021/ja506261t.

(36) Kapuria, N.; Patil, N. N.; Sankaran, A.; Laffir, F.; Geaney, H.; Magner, E.; Scanlon, M.; Ryan, K. M.; Singh, S. Engineering polymorphs in colloidal metal dichalcogenides: precursor-mediated phase control, molecular insights into crystallisation kinetics and promising electrochemical activity. *Journal of Materials Chemistry A* **2023**, *11* (21), 11341-11353, 10.1039/D2TA09892J. DOI: 10.1039/D2TA09892J.

(37) Pippia, G.; Van Hamme, D.; Martín-García, B.; Prato, M.; Moreels, I. A colloidal route to semiconducting tungsten disulfide nanosheets with monolayer thickness. *Nanoscale* **2022**, *14* (42), 15859-15868, 10.1039/D2NR04307F. DOI: 10.1039/D2NR04307F.

(38) Niebur, A.; Söll, A.; Haizmann, P.; Strolka, O.; Rudolph, D.; Tran, K.; Renz, F.; Frauendorf, A. P.; Hübner, J.; Peisert, H.; et al. Untangling the intertwined: metallic to semiconducting phase transition of colloidal $MoS_2$ nanoplatelets and nanosheets. *Nanoscale* **2023**, *15* (12), 5679-5688, 10.1039/D3NR00096F. DOI: 10.1039/D3NR00096F.

(39) Fröhlich, M.; Kögel, M.; Hiller, J.; Kahlmeyer, L.; Meixner, A. J.; Scheele, M.; Meyer, J. C.; Lauth, J. Colloidal 2D $Mo_{1-x}W_xS_2$ nanosheets: an atomic- to ensemble-level spectroscopic study. *Physical Chemistry Chemical Physics* **2024**, *26* (17), 13271-13278, 10.1039/D4CP00530A. DOI: 10.1039/D4CP00530A.

(40) Mutyala, C. S.; Pippia, G.; Tanghe, I.; Martín-García, B.; Rousaki, A.; Vandenabeele, P.; Schiettecatte, P.; Moreels, I.; Geiregat, P. Charge Carrier Dynamics in Colloidally Synthesized Monolayer $MoX_2$ Nanosheets. *The Journal of Physical Chemistry Letters* **2023**, *14* (10), 2620-2626. DOI: 10.1021/acs.jpclett.3c00278.

(41) Schiettecatte, P.; Singh, S.; Zhou, P.; Hens, Z. The Dynamic Interaction of Surfactants with Colloidal Molybdenum Disulfide Nanosheets Calls for Thermodynamic Stabilization by Solvents. *Langmuir* **2023**, *39* (18), 6568-6579. DOI: 10.1021/acs.langmuir.3c00546.

(42) Kumar, S.; Fröhlich, M.; Velja, S.; Kögel, M.; Strolka, O.; Niebur, A.; Ginzburg, S.; Ramzan, M. S.; Meyer, J. C.; Lauth, J.; et al. Atomistic Origin of Photoluminescence Quenching in Colloidal $MoS_2$ and $WS_2$ Nanoplatelets. *Nano Letters* **2026**, *26* (9), 3064-3072. DOI: 10.1021/acs.nanolett.5c05893.

(43) Zhou, M.; Zhang, Z.; Huang, K.; Shi, Z.; Xie, R.; Yang, W. Colloidal preparation and electrocatalytic hydrogen production of MoS2 and WS2 nanosheets with controllable lateral sizes and layer numbers. *Nanoscale* **2016**, *8* (33), 15262-15272, 10.1039/C6NR04775K. DOI: 10.1039/C6NR04775K.

(44) Xemaire, M.; Braun, T.; Kögel, M.; Strolka, O.; Rudolph, D. A.; Meyer, J. C.; Lauth, J. Unraveling the tunable transient optical response of colloidal 2D $Mo_{1-x}W_xS_2$ nanosheets. *Faraday Discussions,* **2026**, *268,* 218 - 231. DOI: 10.1039/d6fd00079g .

(45) Darlington, T. P.; Carmesin, C.; Florian, M.; Yanev, E.; Ajayi, O.; Ardelean, J.; Rhodes, D. A.; Ghiotto, A.; Krayev, A.; Watanabe, K.; et al. Imaging strain-localized excitons in nanoscale bubbles of monolayer $WSe_2$ at room temperature. *Nature Nanotechnology* **2020**, *15* (10), 854-860. DOI: 10.1038/s41565-020-0730-5.

(46) Sie, E. J.; Steinhoff, A.; Gies, C.; Lui, C. H.; Ma, Q.; Rösner, M.; Schönhoff, G.; Jahnke, F.; Wehling, T. O.; Lee, Y.-H.; et al. Observation of Exciton Redshift–Blueshift Crossover in Monolayer $WS_2$. *Nano Letters* **2017**, *17* (7), 4210-4216. DOI: 10.1021/acs.nanolett.7b01034.

(47) Backes, C.; Smith, R. J.; McEvoy, N.; Berner, N. C.; McCloskey, D.; Nerl, H. C.; O'Neill, A.; King, P. J.; Higgins, T.; Hanlon, D.; et al. Edge and confinement effects allow in situ measurement of size and thickness of liquid-exfoliated nanosheets. *Nature Communications* **2014**, *5* (1), 4576. DOI: 10.1038/ncomms5576.

(48) Zipfel, J.; Kulig, M.; Perea-Causín, R.; Brem, S.; Ziegler, J. D.; Rosati, R.; Taniguchi, T.; Watanabe, K.; Glazov, M. M.; Malic, E.; et al. Exciton diffusion in monolayer semiconductors with suppressed disorder. *Physical Review B* **2020**, *101* (11), 115430. DOI: 10.1103/PhysRevB.101.115430.

(49) Li, Q.; Lian, T. Area- and Thickness-Dependent Biexciton Auger Recombination in Colloidal CdSe Nanoplatelets: Breaking the "Universal Volume Scaling Law". *Nano Letters* **2017**, *17* (5), 3152-3158. DOI: 10.1021/acs.nanolett.7b00587.

(50) Plechinger, G.; Nagler, P.; Kraus, J.; Paradiso, N.; Strunk, C.; Schüller, C.; Korn, T. Identification of excitons, trions and biexcitons in single-layer WS2. *physica status solidi (RRL) – Rapid Research Letters* **2015**, *9* (8), 457-461. DOI: https://doi.org/10.1002/pssr.201510224.

(51) Adler, E. R.; Le, T. D.; Boulares, I.; Boyd, R.; He, Y.; Rhodes, D.; Van Keuren, E.; Barbara, P.; Najmaei, S. Observation of Multi-Phonon Emission in Monolayer $WS_2$ on Various Substrates. In *Nanomaterials*, 2024; Vol. 14, p 37.

(52) Demircan, A.; Evlyukhin, A. B.; Morgner, U.; Melchert, O.; Babushkin, I.; Zhang, S.; Biancalana, F.; Mei, C.; Fan, J.; Yang, P.; et al. Roadmap on 5D Photonics for nano and beyond. *Nano Futures* **2026**.

(53) Ding, Y.; Wang, Y.; Ni, J.; Shi, L.; Shi, S.; Tang, W. First principles study of structural, vibrational and electronic properties of graphene-like $MX_2$ (M=Mo, Nb, W, Ta; X=S, Se, Te) monolayers. *Physica B: Condensed Matter* **2011**, *406* (11), 2254-2260. DOI: https://doi.org/10.1016/j.physb.2011.03.044.

(54) Schutte, W. J.; De Boer, J. L.; Jellinek, F. Crystal structures of tungsten disulfide and diselenide. *Journal of Solid State Chemistry* **1987**, *70* (2), 207-209. DOI: 10.1016/0022-4596(87)90057-0.

(55) Frauendorf, A. P.; Niebur, A.; Harms, L.; Shree, S.; Urbaszek, B.; Oestreich, M.; Hübner, J.; Lauth, J. Room Temperature Micro-Photoluminescence Studies of Colloidal $WS_2$ Nanosheets. *The Journal of Physical Chemistry C* **2021**, *125* (34), 18841-18848. DOI: 10.1021/acs.jpcc.1c06240.

(56) Zhao, Y.; Fröhlich, M.; Kögel, M.; Strolka, O.; Niebur, A.; Parker, T.; Schneider, F.; Meixner, A. J.; Meyer, J. C.; Zhang, D.; et al. Second-harmonic generation and photoluminescence properties of colloidal $WS_2$ monolayers deposited from solution. *Nanoscale Horizons* **2025**, *10* (12), 3469-3477, 10.1039/D5NH00515A. DOI: 10.1039/D5NH00515A.

(57) Pippia, G.; Rousaki, A.; Barbone, M.; Billet, J.; Brescia, R.; Polovitsyn, A.; Martín-García, B.; Prato, M.; De Boni, F.; Petrić, M. M.; et al. Colloidal Continuous Injection Synthesis of Fluorescent $MoX_2$ (X = S, Se) Nanosheets as a First Step Toward Photonic Applications. *ACS Applied Nano Materials* **2022**, *5* (8), 10311-10320. DOI: 10.1021/acsanm.2c01470.

(58) McCreary, K. M.; Hanbicki, A. T.; Sivaram, S. V.; Jonker, B. T. A- and B-exciton photoluminescence intensity ratio as a measure of sample quality for transition metal dichalcogenide monolayers. *APL Materials* **2018**, *6* (11). DOI: 10.1063/1.5053699 (acccessed 6/15/2026).

(59) Kłopotowski, Ł.; Backes, C.; Mitioglu, A. A.; Vega-Mayoral, V.; Hanlon, D.; Coleman, J. N.; Ivanov, V. Y.; Maude, D. K.; Plochocka, P. Revealing the nature of excitons in liquid exfoliated monolayer tungsten disulphide. *Nanotechnology* **2016**, *27* (42), 425701. DOI: 10.1088/0957-4484/27/42/425701.

(60) Selig, M.; Berghäuser, G.; Raja, A.; Nagler, P.; Schüller, C.; Heinz, T. F.; Korn, T.; Chernikov, A.; Malic, E.; Knorr, A. Excitonic linewidth and coherence lifetime in monolayer transition metal dichalcogenides. *Nature Communications* **2016**, *7* (1), 13279. DOI: 10.1038/ncomms13279.

(61) Chernikov, A.; Ruppert, C.; Hill, H. M.; Rigosi, A. F.; Heinz, T. F. Population inversion and giant bandgap renormalization in atomically thin $WS_2$ layers. *Nature Photonics* **2015**, *9* (7), 466-470. DOI: 10.1038/nphoton.2015.104.

(62) Orrit, M.; Bernard, J. Single pentacene molecules detected by fluorescence excitation in a p-terphenyl crystal. *Physical Review Letters* **1990**, *65* (21), 2716-2719. DOI: 10.1103/PhysRevLett.65.2716.

(63) Stracke, F.; Blum, C.; Becker, S.; Müllen, K.; Meixner, A. J. Correlation of Emission Intensity and Spectral Diffusion in Room Temperature Single-Molecule Spectroscopy. *ChemPhysChem* **2005**, *6* (7), 1242-1246. DOI: https://doi.org/10.1002/cphc.200400436.

(64) Ambrose, W. P.; Moerner, W. E. Fluorescence spectroscopy and spectral diffusion of single impurity molecules in a crystal. *Nature* **1991**, *349* (6306), 225-227. DOI: 10.1038/349225a0.

(65) Hoogenboom, J. P.; Hernando, J.; van Dijk, E. M. H. P.; van Hulst, N. F.; García-Parajó, M. F. Power-Law Blinking in the Fluorescence of Single Organic Molecules. *ChemPhysChem* **2007**, *8* (6), 823-833. DOI: https://doi.org/10.1002/cphc.200600783.

(66) Dickson, R. M.; Cubitt, A. B.; Tsien, R. Y.; Moerner, W. E. On/off blinking and switching behaviour of single molecules of green fluorescent protein. *Nature* **1997**, *388* (6640), 355-358. DOI: 10.1038/41048.

(67) Chow, P. K.; Jacobs-Gedrim, R. B.; Gao, J.; Lu, T.-M.; Yu, B.; Terrones, H.; Koratkar, N. Defect-Induced Photoluminescence in Monolayer Semiconducting Transition Metal Dichalcogenides. *ACS Nano* **2015**, *9* (2), 1520-1527. DOI: 10.1021/nn5073495.

(68) Backes, C.; Szydłowska, B. M.; Harvey, A.; Yuan, S.; Vega-Mayoral, V.; Davies, B. R.; Zhao, P.-l.; Hanlon, D.; Santos, E. J. G.; Katsnelson, M. I.; et al. Production of Highly Monolayer Enriched Dispersions of Liquid-Exfoliated Nanosheets by Liquid Cascade Centrifugation. *ACS Nano* **2016**, *10* (1), 1589-1601. DOI: 10.1021/acsnano.5b07228.

(69) Bozheyev, F.; Valiev, D.; Nemkayeva, R.; An, V.; Tikhonov, A.; Sugurbekova, G. Pulsed cathodoluminescence of WS2 nanocrystals at various electron excitation energy densities: Defect induced sub-band gap emission. *Journal of Luminescence* **2017**, *192*, 1308-1312. DOI: https://doi.org/10.1016/j.jlumin.2017.09.015.

(70) Sun, H.; Yang, Q.; Wang, J.; Ding, M.; Cheng, M.; Liao, L.; Cai, C.; Chen, Z.; Huang, X.; Wang, Z.; et al. Unveiling sulfur vacancy pairs as bright and stable color centers in monolayer WS2. *Nature Communications* **2024**, *15* (1), 9476. DOI: 10.1038/s41467-024-53880-4.

(71) Brown, R. H.; Twiss, R. Q. Correlation between Photons in two Coherent Beams of Light. *Nature* **1956**, *177* (4497), 27-29. DOI: 10.1038/177027a0.

(72) Kimble, H. J.; Dagenais, M.; Mandel, L. Photon Antibunching in Resonance Fluorescence. *Physical Review Letters* **1977**, *39* (11), 691-695. DOI: 10.1103/PhysRevLett.39.691.

(73) Basché, T.; Moerner, W. E.; Orrit, M.; Talon, H. Photon antibunching in the fluorescence of a single dye molecule trapped in a solid. *Physical Review Letters* **1992**, *69* (10), 1516-1519. DOI: 10.1103/PhysRevLett.69.1516.

(74) Palummo, M.; Bernardi, M.; Grossman, J. C. Exciton Radiative Lifetimes in Two-Dimensional Transition Metal Dichalcogenides. *Nano Letters* **2015**, *15* (5), 2794-2800. DOI: 10.1021/nl503799t.

(75) Amani, M.; Lien, D.-H.; Kiriya, D.; Xiao, J.; Azcatl, A.; Noh, J.; Madhvapathy, S. R.; Addou, R.; Kc, S.; Dubey, M.; et al. Near-unity photoluminescence quantum yield in $MoS_2$. *Science* **2015**, *350* (6264), 1065-1068. DOI: 10.1126/science.aad2114 (acccessed 2026/02/19).

(76) Gaur, A. P. S.; Sahoo, S.; Scott, J. F.; Katiyar, R. S. Electron–Phonon Interaction and Double-Resonance Raman Studies in Monolayer WS2. *The Journal of Physical Chemistry C* **2015**, *119* (9), 5146-5151. DOI: 10.1021/jp512540u.

(77) Varshni, Y. P. Temperature dependence of the energy gap in semiconductors. *Physica* **1967**, *34* (1), 149-154. DOI: https://doi.org/10.1016/0031-8914(67)90062-6.

(78) Żuberek, E.; Majak, M.; Lubczyński, J.; Debus, J.; Watanabe, K.; Taniguchi, T.; Ho, C.-H.; Bryja, L.; Jadczak, J. Upconversion photoluminescence excitation reveals exciton–trion and exciton–biexciton coupling in $hBN/WS_2/hBN$ van der Waals heterostructures. *Scientific Reports* **2022**, *12* (1), 13699. DOI: 10.1038/s41598-022-18104-z.

(79) He, Y.-M.; Clark, G.; Schaibley, J. R.; He, Y.; Chen, M.-C.; Wei, Y.-J.; Ding, X.; Zhang, Q.; Yao, W.; Xu, X.; et al. Single quantum emitters in monolayer semiconductors. *Nature Nanotechnology* **2015**, *10* (6), 497-502. DOI: 10.1038/nnano.2015.75.

(80) Turunen, M.; Brotons-Gisbert, M.; Dai, Y.; Wang, Y.; Scerri, E.; Bonato, C.; Jöns, K. D.; Sun, Z.; Gerardot, B. D. Quantum photonics with layered 2D materials. *Nature Reviews Physics* **2022**, *4* (4), 219-236. DOI: 10.1038/s42254-021-00408-0.

(81) Arora, A.; Koperski, M.; Nogajewski, K.; Marcus, J.; Faugeras, C.; Potemski, M. Excitonic resonances in thin films of WSe2: from monolayer to bulk material. *Nanoscale* **2015**, *7* (23), 10421-10429, 10.1039/C5NR01536G. DOI: 10.1039/C5NR01536G.

# Supporting Information

## Colloidal 2D $WS_2$ nanosheets exhibiting high-purity single photon emission

Markus Xemaire,[a#] Jakob Keck,[a#] Marco Kögel,[b,c] Marc Brecht,[a,d] Jannik C. Meyer,[b,c,e] Alfred J. Meixner,[a,e] Jannika Lauth[a,e,f,g]*.

a. Institute of Physical and Theoretical Chemistry, Eberhard Karls University of Tübingen, Auf der Morgenstelle 18, D-72076 Tübingen, Germany
b. Natural and Medical Sciences Institute (NMI) at the Eberhard Karls University of Tübingen, Markwiesenstraße 55, D-72770 Reutlingen, Germany
c. Institute of Applied Physics, Eberhard Karls University of Tübingen, Auf der Morgenstelle 10, D-72076 Tübingen, Germany
d. Center for Process Analysis and Technology (PA&T), School of Life Sciences and Reutlingen Research Institute (RRI), Reutlingen University, Alte Burgstraße 150, D-72762, Reutlingen, Germany
e. Center for Light-Matter-Interaction, Sensors, and Analytics (LISA+), Eberhard Karls University of Tübingen, Auf der Morgenstelle 15, D-72076 Tübingen, Germany
f. Laboratory for Nano and Quantum Engineering (LNQE), Leibniz University Hannover Schneiderberg 39, D-30167 Hannover
g. Cluster of Excellence PhoenixD (Photonics, Optics, and Engineering – Innovation Across Dimensions), Welfengarten 1A, D-30167 Hannover, Germany

## Methods

*S1: Transmission electron microscopy (TEM):* High resolution TEM images were taken on a JEOL ARM 200F equipped with a CETCOR aberration corrector operating at 80 kV. The samples were prepared by drop-casting the $WS_2$ NS solution (in hexane) onto graphene-coated (by Graphenea Easy Transfer) MEMS heating chips (from DENSsolutions). The dry sample was cleaned subsequently by flushing the TEM-grid with hexane and ethanol. To remove mobile carbon contamination, the sample was heated to 300°C under high vacuum conditions of the electron microscope.
*S2: UV-Vis absorption spectroscopy:* Absorption spectra were acquired on an Agilent Cary 5000 spectrometer equipped with a diffuse reflectance accessory in transmission mode using a quartz cuvette (quartz glass high-performance QS200–2500 nm by Hellma) with a path length of 1 cm. The measurements were conducted in 1,2-dichlorbenzene (o-DCB).
*S3: Low temperature confocal photoluminescence spectroscopy:* A hexane dispersion of $WS_2$ NSs (see S6 below) was precipitated by centrifugation. The NSs were then redispersed in toluene (2 ml) and diluted to $10^{-6}$ times the prepared sample concentration containing 2.5% of the amorphous optical polymer matrix Zeonex®. The resulting polymer solution was spin-coated on a silicon wafer cleaned by sonication in a 2% solution of Hellmanex ™ III cuvette cleaning concentrate in deionized water followed by rinsing with deionized water. At this concentration, single $WS_2$ NSs can be observed as well separated diffraction limited intensity hotspots in the confocal scan images (see Fig. S4). For the spectroscopy of individual NS, the sample was mounted onto the scanning stage of the microscope (Fig. S2 "Sample") in the liquid helium bath cryostat (Janis SVT-200) in our custom-built laser scanning confocal microscope (Fig. S2). The sample was photoexcited with a 520 nm pulsed laser beam (LDH-D-C-520, pulse width <100 ps, repetition rates of 40 MHz for TCSPC and 80 MHz for antibunching) focused through a 0.85 NA objective lens directly in front of the sample. Sample positioning, scanning, and focusing were achieved remotely by piezo-controlled low-temperature stages (Attocube, Controller: ANC350, Coarse positioning: ANPx311, ANPz101, Scanning: ANSxy100). The emitted light was collected by the same lens and guided through a 70/30 beam splitter followed by a 580 nm long-pass filter to remove the excitation light. The emitted light is then reflected to the TCSPC/antibunching detection unit (an Andor CCD camera cooled electronically to -70°C) or the spectrometer. The Andor Si500 spectrograph with a 200 l/mm blaze 720 nm grating reaches a 0.806 nm spectral resolution resulting in a minimal linewidth of 2.87 meV at 2.08 eV (position of the A exciton at 5 K).
For measuring the emission dynamics and antibunching, the light was split 50/50 by a non-polarizing beamsplitter onto two micro photon device avalanche photodiodes (APDs). The instrument response function (IRF) was measured in reflection after passing a neutral density (ND) filter for laser attenuation. Data analysis was performed using the PicoQuant Symphotime software by fitting the bi-exponential decay function to the data histograms in the bi-exponential "reconvolution mode" and with the antibunching analysis function.
*S4: Chemicals:* 1,2-Dichlorobenzene (o-DCB, 99%), diethyl ether ($Et_2O$, 99.5%), ethanol (99.8%), n-hexane (97%), oleylamine (OlAm, 98%) and sulfur (99.98%) were purchased from Sigma Aldrich. Tungsten(VI) chloride (99%) was purchased from Alfa Aesar and 1,1,1,3,3,3-hexamethyldisilazane (HMDS, 98%) was purchased from Acros Organics. OlAm was purified as stated below, before being stored under inert gas atmosphere inside a nitrogen filled glovebox. $WCl_6$, sulfur and HMDS were also stored inside a glovebox. All chemicals other than OlAm were used as supplied without further purification.
*S5: OlAm purification:* Following the procedure of Baranov *et al.*,[1] OlAm (100 ml, yellow liquid) in $Et_2O$ (300 ml) was carefully protonated by the addition of concentrated (37%) HCl solution (32 ml). By the addition of acetonitrile (1000 ml, at 0°C) a white waxy solid precipitated. The solution was rigorously stirred in an ice bath for 1h. After discarding the solvent, the oleylammonium chloride was dispersed in $Et_2O$ and neutralized in a separating funnel with a NaOH solution (24.6 g in 300 ml water). The yellow organic phase was washed five times with water (100 ml) before being dried over $Na_2SO_4$. The $Et_2O$ was removed with a rotary evaporator. The OlAm was then vacuum distilled at 200°C over a 2 $cm^3$ cube of Na to yield a colorless liquid.
*S6: Synthesis of $WS_2$ nanosheets:* Inside a glovebox, a three-neck flask was filled with elemental sulfur (2 mg, 62 μmol) in OlAm (20 ml) and HMDS (0.6 ml, 3 mmol). Alongside, the tungsten precursor was prepared by dissolving $WCl_6$ (5 mg, 13 μmol) in OlAm (10 mL). By using a Schlenk apparatus, the reaction vessel containing the sulfur precursor was heated to 320°C under inert atmosphere, before the metal precursor was injected by a syringe pump (10 ml/h). Afterwards, the crude reaction mixture was stirred for one additional hour and then cooled to room temperature. The product was precipitated by the addition of ethanol (60 ml) and washed three times by redispersion in hexane and precipitation by centrifugation at 4000 rpm. Under ambient condition the final $WS_2$ NSs were stored separated in equal parts, one in hexane (2 ml) for easier deposition and the second part in o-DCB (2 ml) for maximum colloidal stability.

*S7 Photon correlation analysis:* Photon coincidences were recorded in a HBT configuration.[2] The emission field was split by a 50/50 Beamsplitter (BS2 in Fig. S2) onto two APDs and the arrival times were stored in the time-tagged mode (T2, HydraHarp 400 without Laser sync) $g^{(2)}$ correlation was obtained by cross correlating the detectors. The coincidence histogram is an array of equidistant (T=12.5 ns at 80 MHz excitation) peaks. Coincidences were evaluated over $\pm 400$ periods folded into one period. Antibunching is seen as the suppression of the central peak compared to the side peaks.[3] This suppression of the central peak is weakened by uncorrelated background (adding a flat baseline) and the peak overlap $L$ due to non fully relaxed excitation at the 80 MHz repetition rate, so the central cell also contains the tails of its two neighbors. Both enter the raw ratio through a single factor, $g^{(2)}_{corr}(0) = 1 - \left[1 - g^{(2)}_{raw}(0)\right]/[\rho(1-L)]$, where $\rho$ is the signal to background ratio.[4] The uncorrected coincidence $g^{(2)}_{raw} = N_c/A_{side}$ is the ratio between the central $-1/T \leq t \leq 1/T$ peak counts $N_c$ and the mean counts per side peak $A_{side}$. The corrected formula reduces to a standard background correction[5] when the emitter fully relaxes between the pulses (L→0). $\rho$ and $L$ are obtained from a joint Poisson-maximum-likelihood fit[6] of the folded side peak and the central peak with shared peak shape, a two-sided exponential with lifetime τ, convolved with a Gaussian, and a flat baseline. In the central cell, neighbor tails are added carrying the side-peak amplitude.

Error Estimate $g^{(2)}$:

The error is given as the $1\sigma$ ($\alpha = 0.3173$, confidence level 68%) Garwood interval as the Gaussian estimate via $\sqrt{N_c}$ is invalid at low central peak counts $N_c$.

$$\left[g^{(2)}_{low}, g^{(2)}_{high}\right] = \left[\frac{\mu_{low}, \mu_{high}}{A_{side}}\right] \text{ with } \mu_{low} = \frac{1}{2}\chi^2_{2N_c,\alpha/2}, \mu_{high} = \frac{1}{2}\chi^2_{2N_c+1,1-\alpha/2}$$

defined as half the $\alpha/2$ quantile of a chi-squared distribution with $2N_c$ degrees of freedom (low) and the $1 - \alpha/2$ quantile of a chi-squared distribution with $2N_c + 2$ degrees of freedom (high). Propagating $g^{(2)}_{low,high}$ through the correction formula gives the stated error estimate.

## Supplementary Figures

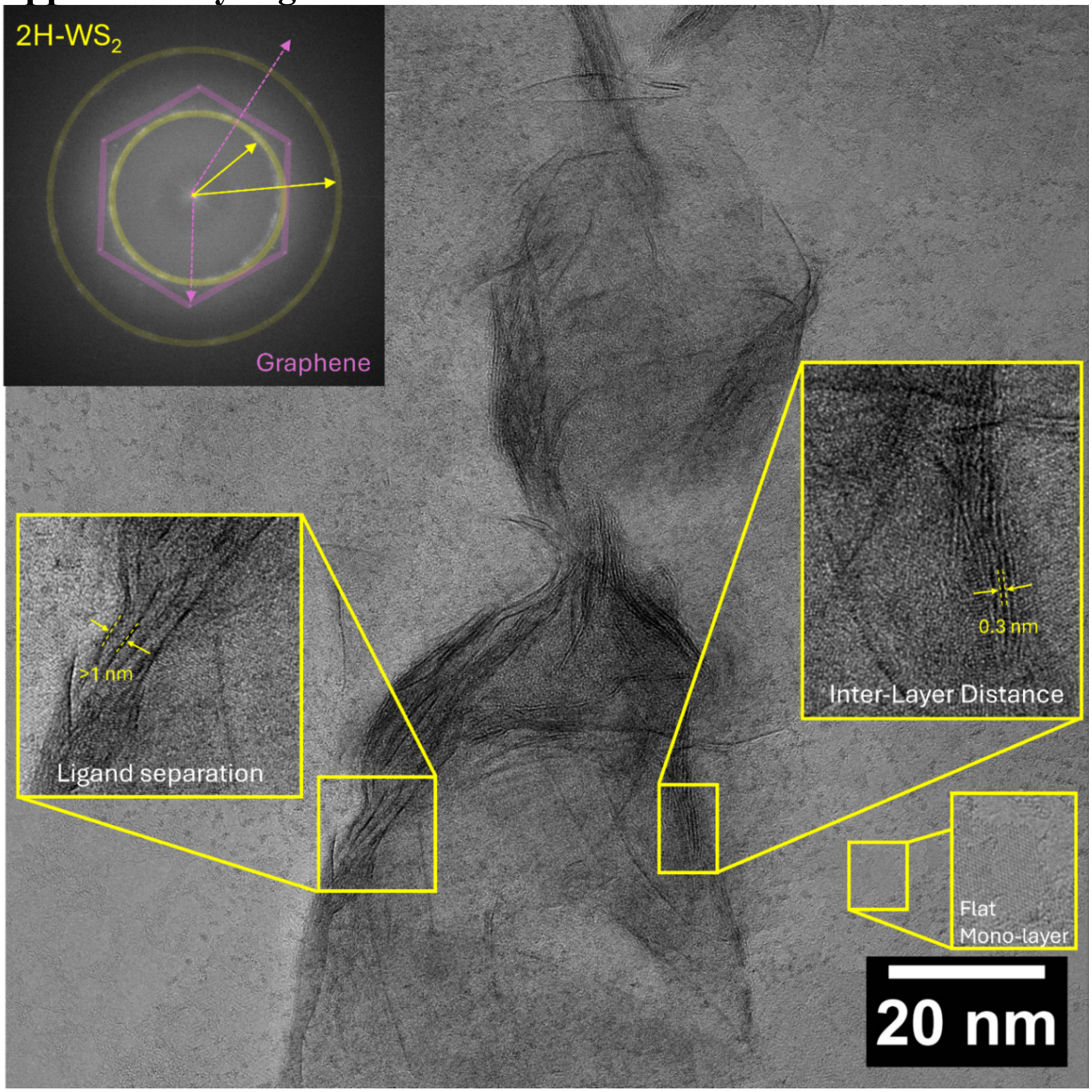

Fig. S2. Overview TEM image at high concentration of $WS_2$, showing agglomeration of NSs. Flat and ligand separated monolayers (varying inter-layer distances >1 nm) as well as few-layers (0.3 nm inter-layer distance)[7] are observed. The FFT inset shows a single orientation for the underlying graphene layer of the TEM grid, while multiple orientations of $WS_2$ NSs yield a circular pattern.

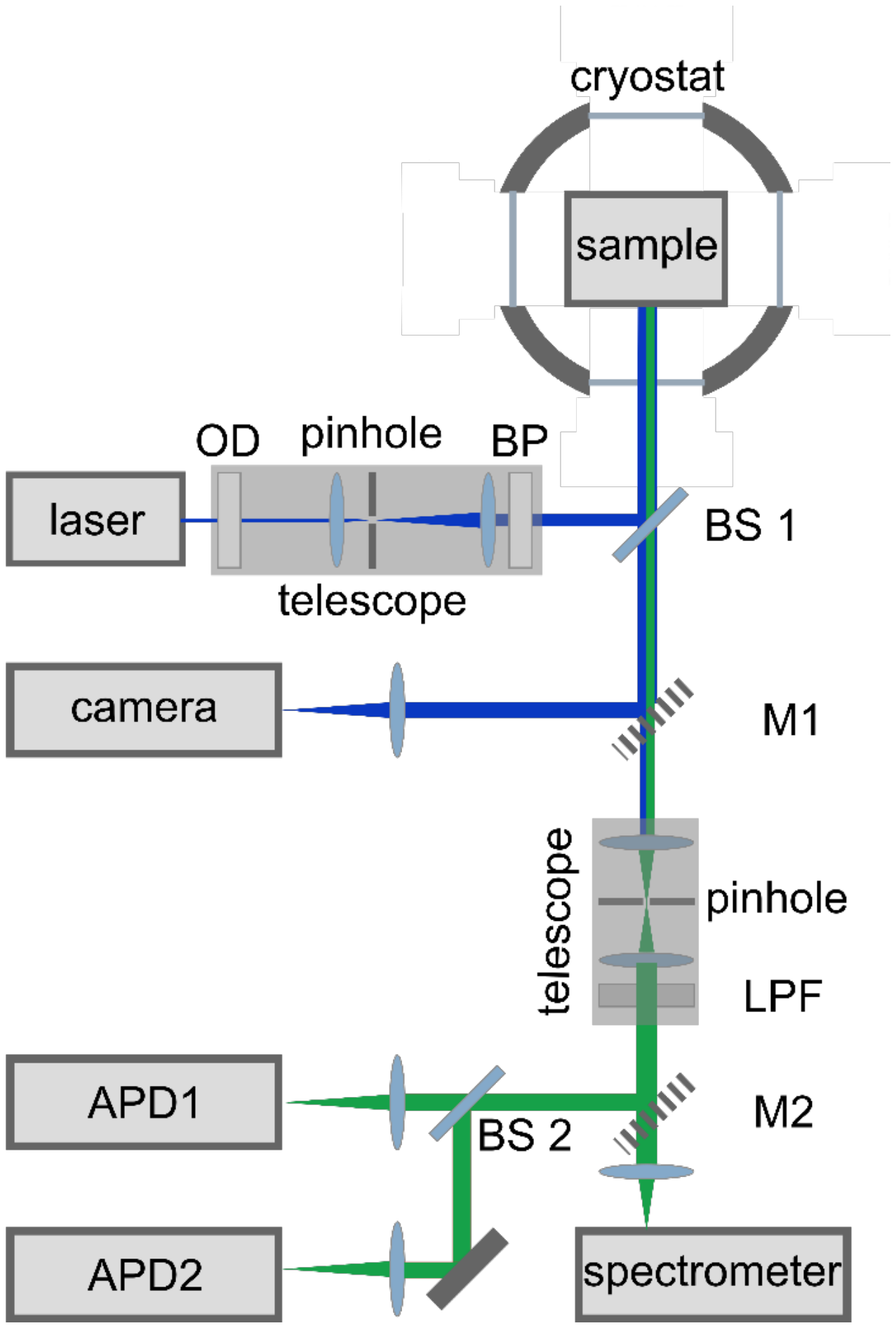

Fig. S3. Experimental setup: The excitation laser (520 nm) is attenuated by a neutral density filter and spatially cleaned by a pinhole positioned in a non-expanding telescope. A band pass filter is positioned after the telescope to remove laser side bands. A 70/30 beamsplitter (BS 1) reflects 30% of the excitation light onto the sample inside the cryostat. The emission light is transmitted with 70% through the BS. The reflected light is projected by a flip mirror (M1) on the camera to observe the laser focus on the sample. To suppress out of focus excitation light, a pinhole is positioned in the non-expanding detection telescope followed by a 580 nm long pass filter (LPF). A flip mirror (M2) selects the detection channel (time-resolved single-photon detector or spectrometer). For TCSPC and antibunching measurements, the emission light is guided onto a 50/50 beamsplitter (BS 2) and projected onto both APDs.

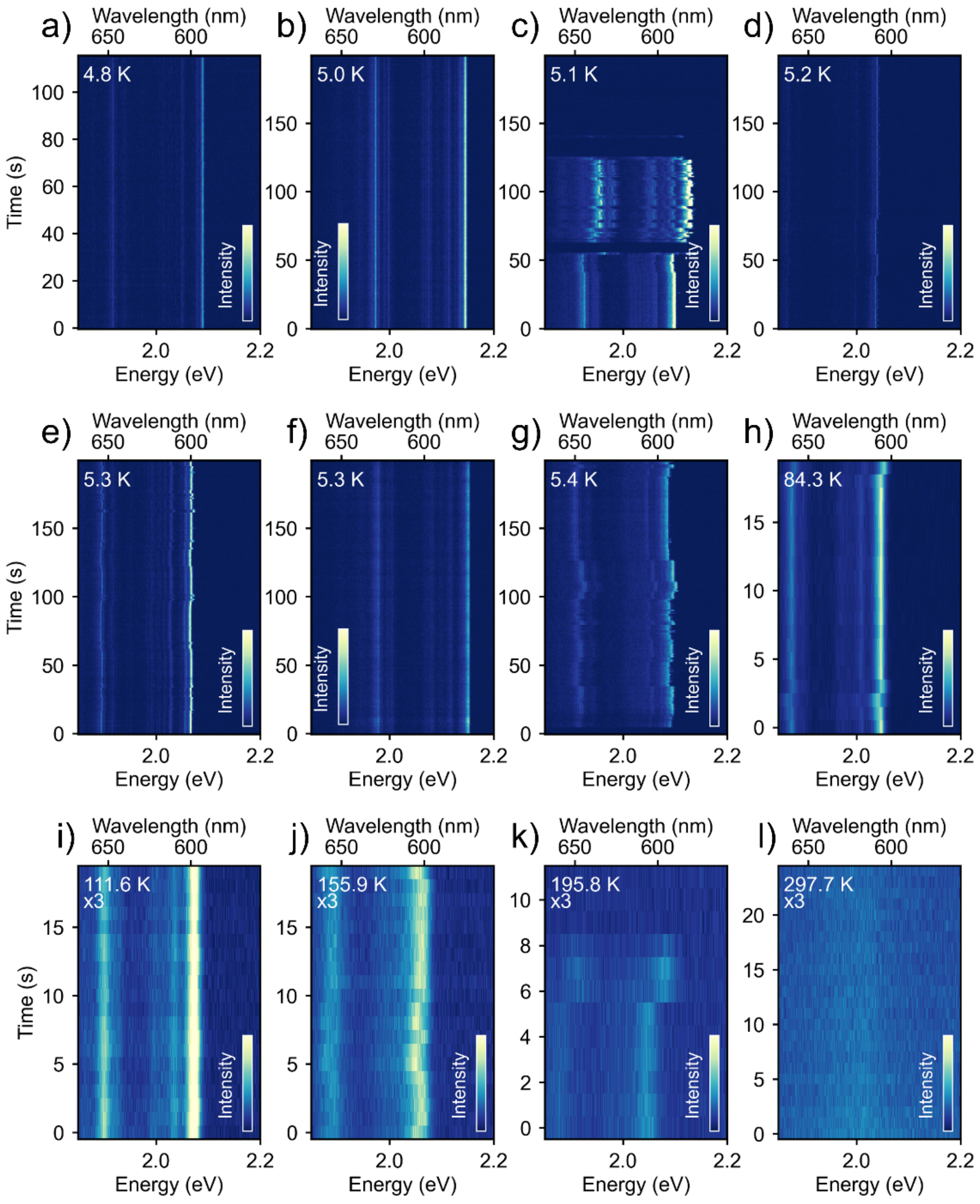

Fig. S3) Sequences of spectra from single $WS_2$ NSs visualized on the same intensity scale and measured at different temperatures, a-g) at ~5 K, the PL of individual NSs slightly differs in spectral position, spectral diffusion behavior and in one case blinking, h-l) with increasing the temperature, broadening, red-shifting and reduced overall intensity (i-l are multiplied by 3 ×) of the PL signal in single $WS_2$ NSs is observed. Photo-damage of single $WS_2$ NSs usually occurs faster at higher temperatures (shorter series < 30 s above 5 K).

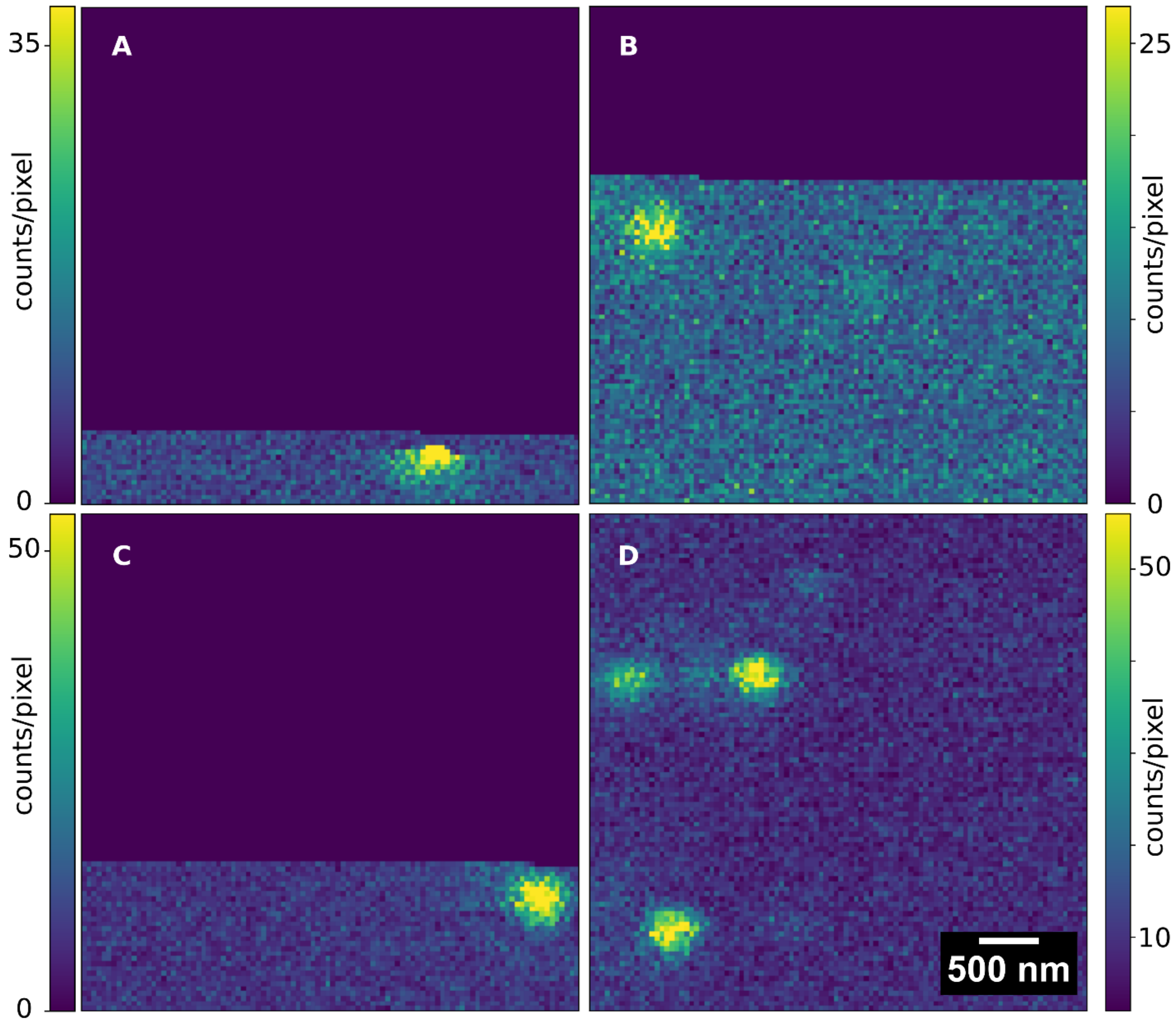

Fig. S4: Confocal laser scan images at 5 K of the well separated NSs. The x-y dimension of the scan (scale bar bottom right) is estimated via the resolution limit due to the temperature dependence of the scan range. The intensity hot spots were fitted with a 2D Gaussian assuming $FWHM = 0.51\,\lambda/NA$ with $\lambda = 595\,nm$ (diffraction limit with pinhole at one airy unit). In A)-C) the scan was stopped close to the nanosheet for following data acquisition.

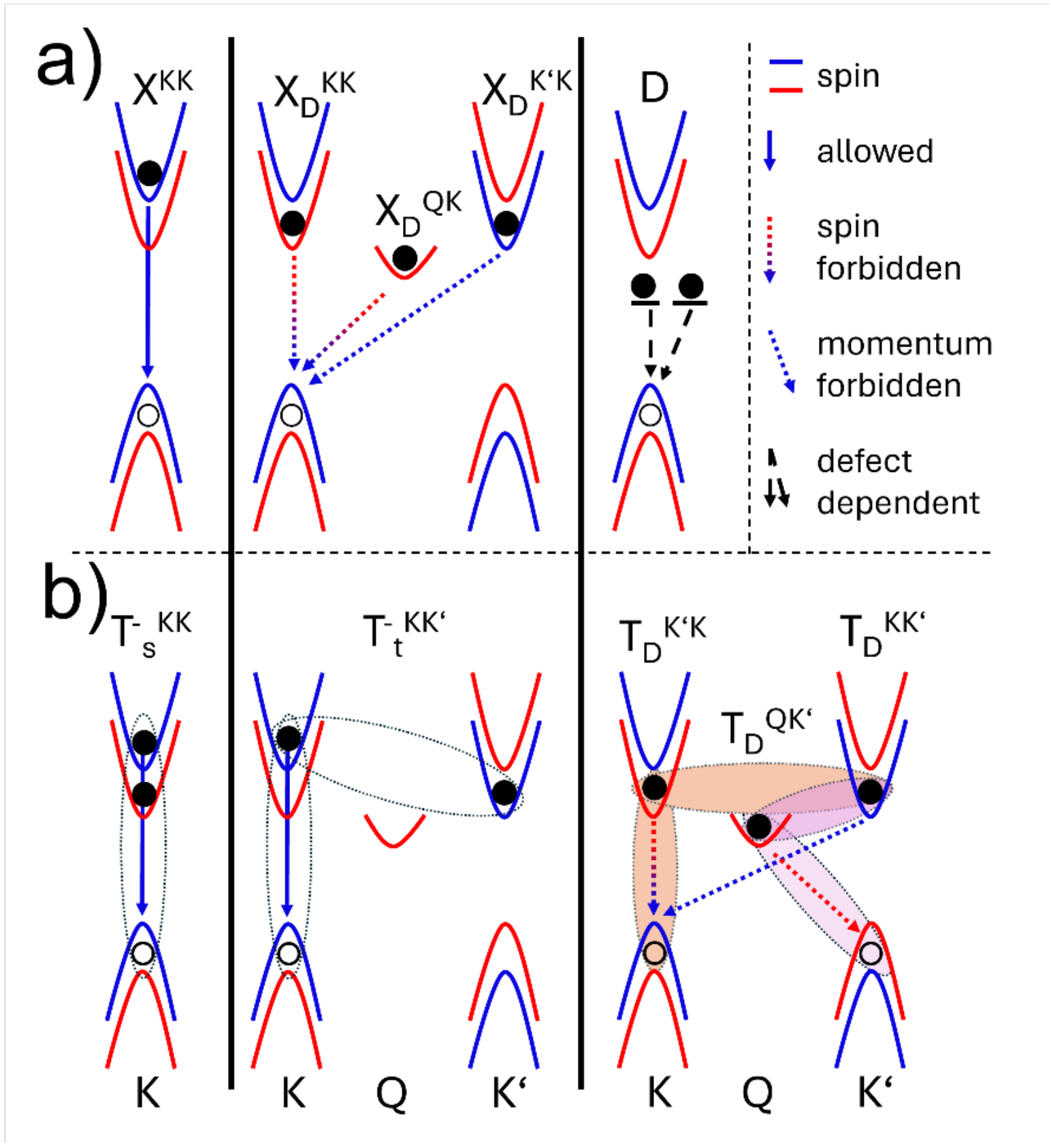

Fig. S5) Summary of possible a) bright ($X^{KK}$), dark ($X_D^{KK}$, $X_D^{QK}$, $X_D^{K'K}$) and defect localized (D) excitons and b) singlet ($T_{-s}^{KK}$), triplet ($T_{-t}^{KK'}$) and dark ($T_D^{K'K}$, $T_D^{QK'}$, $T_D^{KK'}$) trions reported for $WS_2$ at 4 and 78 K.[8]

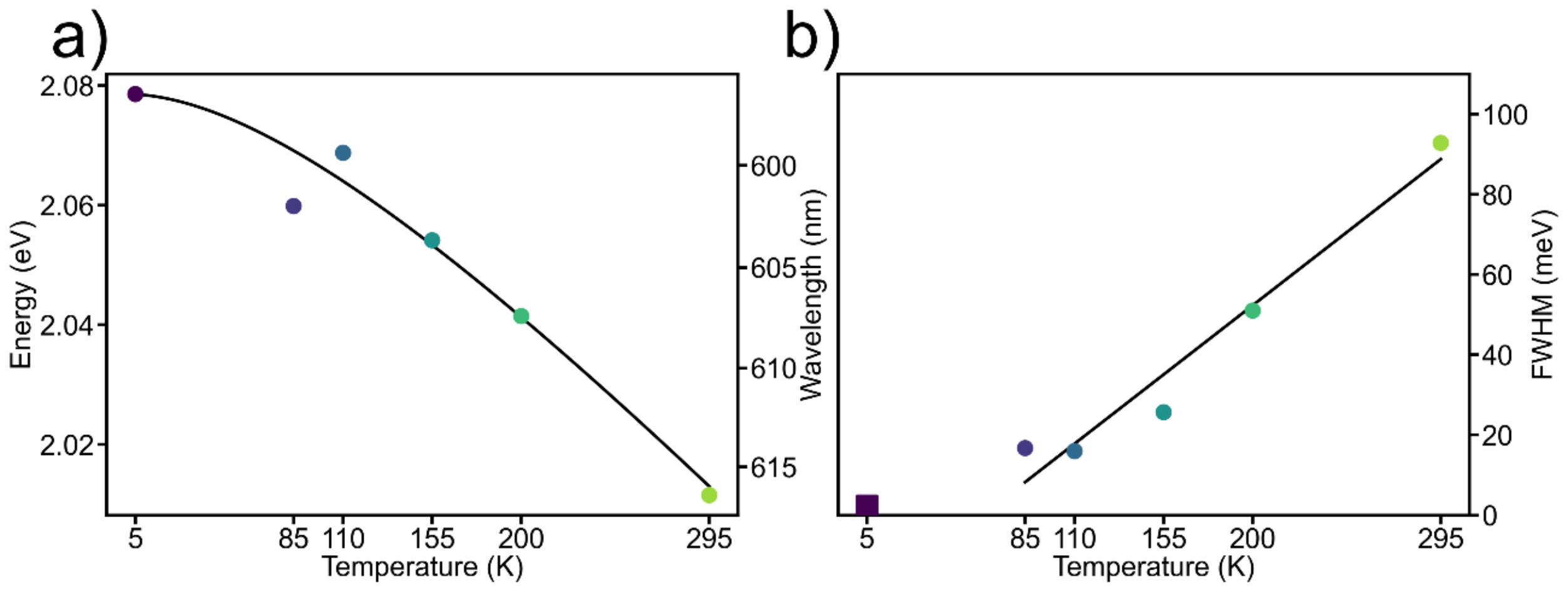

Fig. S6) Median trends of peak position $X_A$ a) and FWHM b) with increasing temperature the peak positions follow the Varshni equation ($\alpha = 3.8 \times 10^{-4} eV/K, \beta = 200\ K$), while broadening between 85 and 295 K is roughly linear ($4 \times 10^{-4} eV/K$). FWHM at 5 K is excluded from the trend as the linewidth is limited by the spectral resolution of the setup.

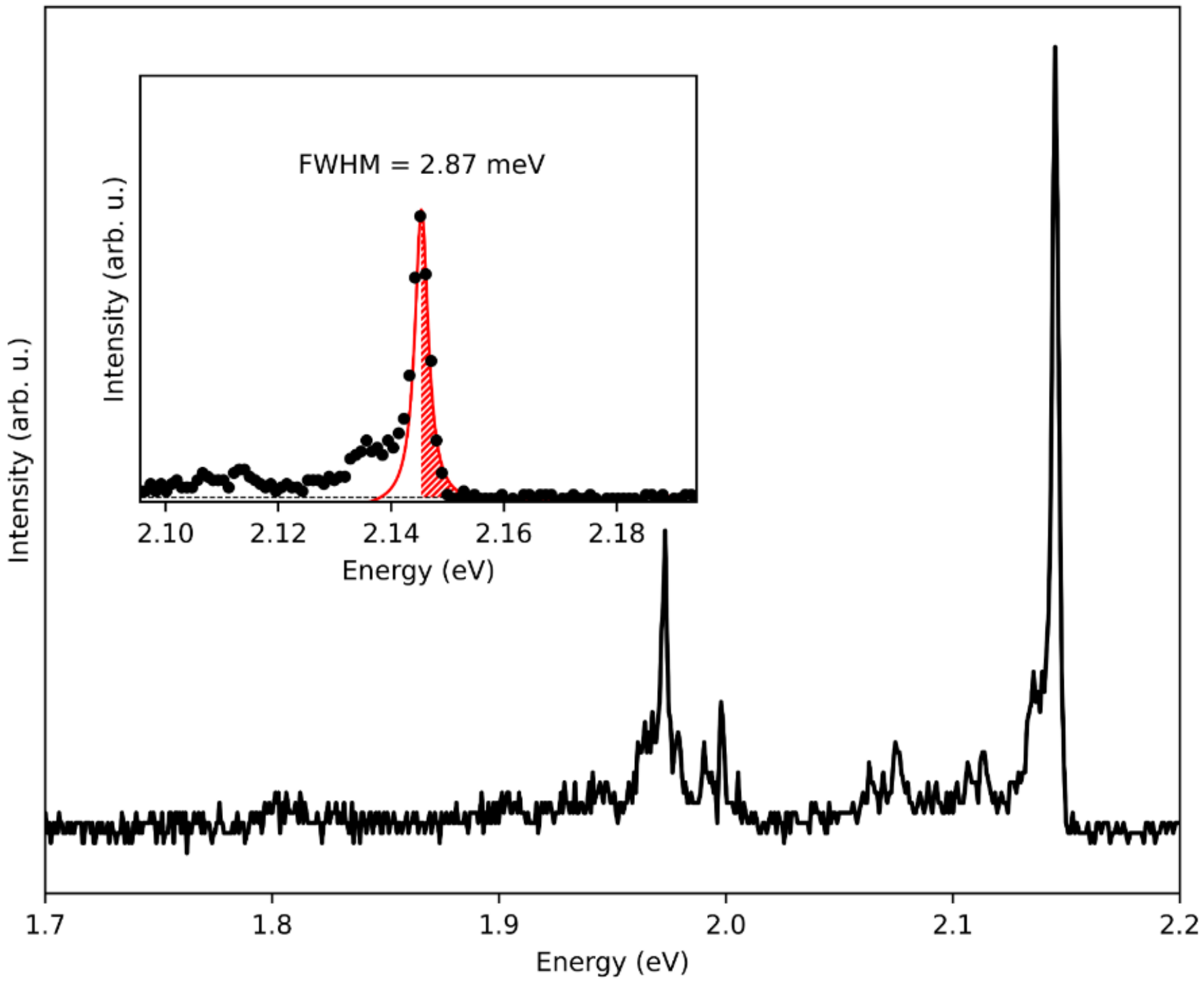

Fig. S7) Full width at half maximum (FWHM) of the neutral excitonic peak $X_A$ at 5 K estimated by Lorentz-fitting the higher energy side of the peak (no shoulder on this side). The found line width exactly matches the spectral resolution of the spectrometer calculated above (Methods S3). Linewidths measurements performed close to 5 K are therefore not included in the trend observed in Fig S6.

**Corresponding Author**

jannika.lauth@uni-tuebingen.de

REFERENCES

[1] Baranov, D.; Lynch, M. J.; Curtis, A. C.; Carollo, A. R.; Douglass, C. R.; Mateo-Tejada, A. M.; Jonas, D. M. *Chem. Mater.* **2019**, *31* (4), 1223-1230.

[2] Brown, R. H.; Twiss, R. Q. *Nature* **1956**, *177* (4497), 27-29.

[3] Kimble, H. J.; Dagenais, M.; Mandel, L. *Phys. Rev. Lett.* **1977**, *39* (11), 691-695.

[4] Beveratos, A.; Brouri, R.; Gacoin, T.; Poizat, J.-P.; Grangier, P. *Phys. Rev. A* **2001**, *64* (6), 061802.

[5] Brouri, R.; Beveratos, A.; Poizat, J.-P.; Grangier, P. *Opt. Lett.* **2000**, *25* (17), 1294-1296.

[6] Baker, S.; Cousins, R. D. *Nucl. Instrum. Methods Phy. Res.* **1984**, *221* (2), 437-442.

[7] Schutte, W. J.; De Boer, J. L.; Jellinek, F. *J.Solid State Chem.* **1987**, *70* (2), 207-209.

[8] Adler, E. R.; Le, T. D.; Boulares, I.; Boyd, R.; He, Y.; Rhodes, D.; Van Keuren, E.; Barbara, P.; Najmaei, S. *Nanomaterials*, **2024**, 14, 37.